\documentclass[english,showpacs,preprint,superscriptaddress]{revtex4-1}
\usepackage{lmodern}
\usepackage{lmodern}
\usepackage[T1]{fontenc}
\usepackage{amsmath}
\usepackage{amssymb}
\usepackage{graphicx}
\usepackage{esint}

\makeatletter

\usepackage{amsthm}\usepackage{latexsym}\usepackage{bm}\usepackage{amsfonts}

\usepackage{babel}

\@ifundefined{showcaptionsetup}{}{%
 \PassOptionsToPackage{caption=false}{subfig}}
\usepackage{subfig}
\makeatother

\usepackage{babel}
\begin{document}

\title{Explicit high-order non-canonical symplectic particle-in-cell algorithms
for Vlasov-Maxwell systems}

\author{Jianyuan Xiao}

\affiliation{School of Nuclear Science and Technology and Department of Modern
Physics, University of Science and Technology of China, Hefei, Anhui
230026, China}

\affiliation{Key Laboratory of Geospace Environment, CAS, Hefei, Anhui 230026,
China}

\author{Hong Qin }

\email{corresponding author: hongqin@ustc.edu.cn}

\affiliation{School of Nuclear Science and Technology and Department of Modern
Physics, University of Science and Technology of China, Hefei, Anhui
230026, China}

\affiliation{Plasma Physics Laboratory, Princeton University, Princeton, NJ 08543}

\author{Jian Liu}

\affiliation{School of Nuclear Science and Technology and Department of Modern
Physics, University of Science and Technology of China, Hefei, Anhui
230026, China}

\affiliation{Key Laboratory of Geospace Environment, CAS, Hefei, Anhui 230026,
China}

\author{Yang He}

\affiliation{School of Nuclear Science and Technology and Department of Modern
Physics, University of Science and Technology of China, Hefei, Anhui
230026, China}

\affiliation{Key Laboratory of Geospace Environment, CAS, Hefei, Anhui 230026,
China}

\author{Ruili Zhang}

\affiliation{School of Nuclear Science and Technology and Department of Modern
Physics, University of Science and Technology of China, Hefei, Anhui
230026, China}

\affiliation{Key Laboratory of Geospace Environment, CAS, Hefei, Anhui 230026,
China}

\author{Yajuan Sun}

\affiliation{LSEC, Academy of Mathematics and Systems Science, Chinese Academy
of Sciences, P.O. Box 2719, Beijing 100190, China }
\begin{abstract}
Explicit high-order non-canonical symplectic particle-in-cell algorithms
for classical particle-field systems governed by the Vlasov-Maxwell
equations are developed. The algorithm conserves a discrete non-canonical
symplectic structure derived from the Lagrangian of the particle-field
system, which is naturally discrete in particles. The electromagnetic
field is spatially-discretized using the method of discrete exterior
calculus with high-order interpolating differential forms for a cubic
grid. The resulting time-domain Lagrangian assumes a non-canonical
symplectic structure. It is also gauge invariant and conserves charge.
The system is then solved using a structure-preserving splitting method
discovered by He et al., which produces five exactly-soluble sub-systems,
and high-order structure-preserving algorithms follow by combinations.
The explicit, high-order, and conservative nature of the algorithms
is especially suitable for long-term simulations of particle-field
systems with extremely large number of degrees of freedom on massively
parallel supercomputers. The algorithms have been tested and verified
by the two physics problems, i.e., the nonlinear Landau damping and
the electron Bernstein wave.
\end{abstract}

\keywords{explicit method, non-canonical structure, symplectic algorithm}

\pacs{52.65.Rr, 52.25.Dg}

\maketitle
\global\long\def\EXP{\times10}
 \global\long\def\rmd{\mathrm{d}}
 \global\long\def\xs{ \mathbf{x}_{s}}
 \global\long\def\dotxs{\dot{\mathbf{x}}_{s}}
 \global\long\def\bfx{\mathbf{x}}
 \global\long\def\bfv{\mathbf{v}}
 \global\long\def\bfA{\mathbf{A}}
 \global\long\def\bfB{\mathbf{B}}
 \global\long\def\bfE{\mathbf{E}}
 \global\long\def\bfu{\mathbf{u}}
 \global\long\def\bfe{\mathbf{e}}
 \global\long\def\bfd{\mathbf{d}}
 \global\long\def\rme{\mathrm{e}}
 \global\long\def\rmi{\mathrm{i}}
 \global\long\def\rmq{\mathrm{q}}
 \global\long\def\ope{\omega_{pe}}
 \global\long\def\oce{\omega_{ce}}
 \global\long\def\REF#1{Ref.~\cite{#1}}
 \global\long\def\CURLD{ {\mathrm{curl_{d}}}}
 \global\long\def\DIVD{ {\mathrm{div_{d}}}}
 \global\long\def\CURLDP{ {\mathrm{curl_{d}}'}}
 \global\long\def\cpt{\captionsetup{justification=raggedright }}
 \global\long\def\act{\mathcal{A}}

\section{Introduction}

The importance of numerical solutions for the Vlasov-Maxwell (VM)
system cannot be overemphasized. In most cases, important and interesting
characteristics of the VM system are the long-term behaviors and multi-scale
structures, which demand long-term accuracy and fidelity of numerical
calculations. Conventional algorithms for the VM systems used in general
do not preserve the geometric structures of the physical systems,
such as the local energy-momentum conservation law and the symplectic
structure. For these algorithms, the truncation errors are small only
for each time-step. For example, the truncation error of a fourth
order Runge-Kutta method is of the fifth order of the step-size for
each time-step. However, numerical errors from different time-steps
accumulate coherently with time and long-term simulation results are
not reliable. To overcome this difficult, a series of geometric algorithms,
which preserve the geometric structures of theoretical models in plasma
physics have been developed recently. 

At the single particle level, canonical Hamilton equation for charged
particle dynamics can be integrated using the standard canonical symplectic
integrators developed in the late 1980s \cite{Ruth83,Feng85,Feng86,Feng10,Forest90,Channell90,Candy91,marsden2001discrete,Hairer02}.
Since the Hamiltonian expressed in terms of the canonical momentum
is not separable, it is believed that symplectic algorithms applicable
are in general implicit. Recent studies show that this is not the
case, and high order explicit symplectic algorithms for charged particle
dynamics have been discovered \cite{chin2008symplectic,he2015hamiltonian,he2015explicit}.

For the most-studied guiding center dynamics in magnetized plasmas,
a non-canonical variational symplectic integrator has been developed
and applied \cite{PhysRevLett.100.035006,qin2009variational,li2011variational,squire2012geometric,kraus2013variational,zhang2014canonicalization,ellison2015development}.
It is also recently discovered that the popular Boris algorithm is
actually a volume-preserving algorithm \cite{qin2013boris}. This
revelation stimulated new research activities \cite{zhang2015comment,ellison2015comment}.
For example, high-order volume-preserving methods \cite{he2015volume}
and relativistic volume-preserving methods \cite{zhang2015volume}
have been worked out systematically. 

For collective dynamics of the particle-field system governed by the
Vlasov-Maxwell equations \cite{qin2007geometric,qin2014field}, Squire
et al. \cite{Squire4748,squire2012gauge} constructed the first geometric,
structure-preserving algorithm by discretizing a geometric variational
principle \cite{qin2007geometric}. It has been applied in simulation
studies of nonlinear radio-frequency waves in magnetized plasmas \cite{xiao2013variational,xiao2015variational}.
Similar methods apply to Vlasov-Poisson system as well \cite{evstatiev2013variational,evstatiev2014application,Shadwick14}.
We can also discretize directly the Poisson structures of the Vlasov-Maxwell
system. A canonical symplectic Particle-in-Cell (PIC) algorithm is
found by discretizing the canonical Poisson bracket \cite{qin2015canonical},
and non-canonical symplectic methods are being developed using the
powerful Hamiltonian splitting technique \cite{crouseilles2015hamiltonian,Qin15JCP,he2015hamiltonian}
that preserve the non-canonical Morrison-Marsden-Weinstein bracket
\cite{Morrison80,Weinstein81,marsden1982hamiltonian,burby2014hamiltonian}
for the VM equations. Of course, geometric structure-preserving algorithms
are expected for reduced systems as well. For example, an structure-preserving
algorithm has been developed for ideal MHD equations \cite{zhou2014variational},
and applied to study current sheet formation in an ideal plasma without
resistivity \cite{zhou2015formation}. The superiority of these geometric
algorithms has been demonstrated. This should not be surprising because
geometric algorithms are built on the more fundamental field-theoretical
formalism, and are directly linked to the perfect form, i.e., the
variational principle of physics. The fact that the most elegant form
of theory is also the most effective algorithm is philosophically
satisfactory. 

In this paper, we present an explicit, high order, non-canonical symplectic
PIC algorithm for the Vlasov-Maxwell system. The algorithm conserves
a discrete non-canonical symplectic structure derived from the Lagrangian
of the particle-field system \cite{qin2007geometric,qin2014field}.
The Lagrangian is naturally discrete in particles, and the electromagnetic
field is discretized using the method of discrete exterior calculus
(DEC). An important technique for interpolating differential forms
over several grid cells are developed, which generalizes the construction
of Whitney forms to higher orders. The resulting Lagrangian is continuous
in time and assumes a non-canonical symplectic structure, the dimension
of which is finite but large. Because the electromagnetic field is
interpolated as differential forms, the time-domain Lagrangian is
also gauge invariant and conserves charge. From this Lagrangian, we
can readily derive the non-canonical symplectic structure for the
dynamics, and the system is solved using a splitting method discovered
by He et al. \cite{he2015hamiltonian,he2015explicit}. The splitting
produces five exactly-soluble sub-systems, and high-order structure-preserving
algorithms follow by combinations. We note that for previous symplectic
PIC methods \cite{Squire4748,squire2012gauge,xiao2013variational,xiao2015variational,qin2015canonical},
high-order algorithms are implicit in general, and only the first
order method can be made explicit \cite{qin2015canonical}. The explicit
and high-order nature of the symplectic algorithms developed in the
present study made it especially suitable for long-term simulations
of particle-field systems with extremely large number of degrees of
freedom on massively parallel supercomputers.

The paper is organized as follows. The non-canonical symplectic PIC
algorithm is derived in Sec. II with an appendix on Whitney forms
and their generalization to high orders. In Sec. III, the developed
algorithm is tested and verified by two physics problems, i.e., the
nonlinear Landau damping and the electron Bernstein wave.

\section{Non-Canonical Symplectic Particle-in-Cell Algorithms}

We start from the Lagrangian of a collection of charged particles
and electromagnetic field \cite{qin2007geometric,qin2014field} 
\begin{eqnarray}
L & = & \iiint\rmd\bfx\left(\frac{\epsilon_{0}}{2}\left(-\dot{\bfA}\left(\bfx\right)-\nabla\phi\left(\bfx\right)\right)^{2}-\frac{1}{2\mu_{0}}\left(\nabla\times\bfA\left(\bfx\right)\right)^{2}+\right.\nonumber \\
 &  & \left.\sum_{s}\delta\left(\bfx-\bfx_{s}\right)\left(\frac{1}{2}m_{s}\dot{\bfx}_{s}^{2}+q_{s}\bfA\left(\bfx\right)\cdot\dot{\bfx}_{s}-q_{s}\phi\left(\bfx\right)\right)\right)~,
\end{eqnarray}
where $\bfA\left(\bfx\right)$ and $\phi\left(\bfx\right)$ are the
vector and scalar potentials of the electromagnetic field, $\bfx_{s}$,
$m_{s}$ and $q_{s}$ denote the location, mass and charge of the
$s$-th particle, and $\epsilon_{0}$ and $\mu_{0}$ are the permittivity
and permeability in vacuum. We let $\epsilon_{0}=\mu_{0}=1$ to simplify
the notation.

This Lagrangian is naturally discrete in particles, and we choose
to discretize the electromagnetic field in a cubic mesh. To preserve
the symplectic structure of the system, the method of Discrete Exterior
Calculus (DEC) \cite{hirani2003discrete} is used. The DEC theory
in cubic meshes can be found in Ref. \cite{stern2007geometric}. For
field-particle interaction, the interpolation function is used to
obtain continuous fields from discrete fields. The spatially-discretized
Lagrangian $L_{sd}$ can be written as follows 
\begin{multline}
L_{sd}=\frac{1}{2}\left(\sum_{J}\left(-\dot{\bfA}_{J}-\sum_{I}{\nabla_{\mathrm{d}}}_{JI}\phi_{I}\right)^{2}-\sum_{K}\left(\sum_{J}\CURLD_{KJ}\bfA_{J}\right)^{2}\right)\Delta V+\\
\sum_{s}\left(\frac{1}{2}m_{s}\dot{\bfx}_{s}^{2}+q_{s}\left(\dot{\bfx}_{s}\cdot\sum_{J}W_{\sigma_{1J}}\left(\bfx_{s}\right)\bfA_{J}-\sum_{I}W_{\sigma_{0I}}\left(\bfx_{s}\right)\phi_{I}\right)\right)~,
\end{multline}
where integers $I$, $J$ and $K$ are indices of grid points, and
$\nabla_{\mathrm{d}}$ and $\CURLD$ are the discrete gradient and
curl operators, which are linear operators on the discrete fields
$\bfA_{J}$ and $\phi_{I}$. Functions $W_{\sigma_{0J}}$ and $W_{\sigma_{1I}}$
are interpolation functions for 0-forms (e.g. scalar potential) and
1-forms (e.g. vector potential), respectively. They should be viewed
as maps operating on the discrete 1-form $\bfA_{J}$ and 0-form $\phi_{I}$
to generate continuous forms. More precisely, $W_{\sigma_{1J}}\left(\bfx_{s}\right)\bfA_{J}$
are the components of the continuous 1-form interpolated from the
discrete 1-form $\bfA_{J}$. The idea of form interpolation maps is
due to Whitney, and interpolated forms are called Whitney forms. The
original Whitney forms \cite{whitney1957geometric} are first order
and only for forms in simplicial meshes (e.g. triangle and tetrahedron
meshes). In the present study, we have developed high-order interpolation
maps for a cubic mesh. The details of the construction of $\nabla_{\mathrm{d}}$,
$\CURLD$, $W_{\sigma_{0I}}$, $W_{\sigma_{1J}}$ , and the interpolating
function for 2-forms (e.g. magnetic fields) $W_{\sigma_{2K}}$ are
presented in Appendix \ref{SecWhitneyInterp}. The major new feature
of the form interpolation method adopted here is that the interpolation
for electric field and magnetic field are different. Even for components
in different directions of the same field, the interpolation functions
are not the same. This is very different from traditional cubic interpolations
used in conventional PIC methods \cite{birdsall1991plasma,hockney1988computer,nieter2004vorpal},
where the same interpolation function is used for all components of
electromagnetic fields. The advanced form interpolation method developed
in the present study guarantees the geometric properties of the continuous
system are preserved by the discretized system. 

The action integral is 
\begin{eqnarray}
S=\int\rmd tL_{sd}~,
\end{eqnarray}
and the dynamic equations are obtained from Hamilton's principle,
\begin{eqnarray}
\frac{\delta S}{\delta\bfA_{J}} & = & 0~,\label{EqnMaxwellEqn}\\
\frac{\delta S}{\delta\phi_{I}} & = & 0~,\label{EqnPoisson}\\
\frac{\delta S}{\delta\bfx_{s}} & = & 0~.\label{EqnLorentz}
\end{eqnarray}
Equations (\ref{EqnMaxwellEqn}) and (\ref{EqnPoisson}) are Maxwell's
equations, and Eq. (\ref{EqnLorentz}) is Newton's equation with the
Lorentz force for the $s$-th particle. For the dynamics to be gauge
independent \cite{squire2012geometric}, it requires that the discrete
differential operators and interpolation functions satisfy the following
relations, 
\begin{eqnarray}
\nabla\sum_{I}W_{\sigma_{0I}}\left(\bfx\right)\phi_{I}=\sum_{I,J}W_{\sigma_{1J}}\left(\bfx\right){\nabla_{\mathrm{d}}}_{JI}\phi_{I}~,\\
\nabla\times\sum_{J}W_{\sigma_{1J}}\left(\bfx\right)\bfA_{J}=\sum_{J,K}W_{\sigma_{2K}}\left(\bfx\right){\CURLD}_{KJ}\bfA_{J}~.\label{EqnD1to2FORM}
\end{eqnarray}
The gauge independence of this spatially-discretized system implies
that the dynamics conserves charge automatically. 

Since the dynamics are gauge independent, we can choose any gauge
that is convenient. For simplicity, the temporal gauge, i.e. $\phi_{I}=0$,
is adopted in the present study. To obtain the non-canonical symplectic
structure and Poisson bracket, we look at the Lagrangian 1-form $\gamma$
for the spatially-discretized system defined by $S=\int\gamma$ .
Let $q=[\bfA_{J},\bfx_{s}]$, and the Lagrangian 1-form can be written
as 
\begin{eqnarray}
\gamma=\frac{\partial L_{sd}}{\partial\dot{q}}\bfd q-H\bfd t~,\label{eq:9}
\end{eqnarray}
where $\bfd$ denotes the exterior derivative. In Eq.\,\eqref{eq:9},
\begin{eqnarray}
\frac{\partial L_{sd}}{\partial\dot{q}} & = & [\dot{\bfA}_{J}\Delta V,m_{s}\dot{\bfx}_{s}+q_{s}\sum_{J}W_{\sigma_{1J}}\left(\bfx_{s}\right)\bfA_{J}]~,
\end{eqnarray}
and 
\begin{eqnarray}
H & = & \frac{\partial L_{sd}}{\partial\dot{q}}\dot{q}^{T}-L_{sd}~\\
 & = & \frac{1}{2}\Delta V\left(\sum_{J}\dot{\bfA}_{J}^{2}+\sum_{K}\left(\sum_{J}\CURLD_{KJ}\bfA_{J}\right)^{2}\right)+\sum_{s}\frac{1}{2}m_{s}\dot{\bfx}_{s}^{2}~
\end{eqnarray}
is the Hamiltonian. The dynamical equation of the system can be written
as \cite{QinFields,qin2007geometric} 
\begin{eqnarray}
i_{[\dot{q},\ddot{q},1]}\bfd\gamma=0~,\label{EqnITDGAMMA}
\end{eqnarray}
where $[\dot{q},\ddot{q},1]$ represent vector field $\dot{q}\frac{\partial}{\partial q}+\ddot{q}\frac{\partial}{\partial\dot{q}}+\frac{\partial}{\partial t}.$
The non-canonical symplectic structure is 
\begin{eqnarray}
\Omega=\bfd\left(\frac{\partial L_{sd}}{\partial\dot{q}}\bfd q\right)~,
\end{eqnarray}
and the dynamical equation \eqref{EqnITDGAMMA} is equivalent to 
\begin{eqnarray}
\frac{d}{dt}\left[\begin{array}{c}
\bfA_{J}\\
\bfx_{s}\\
\dot{\bfA}_{J}\\
\dot{\mathbf{x}}_{s}
\end{array}\right] & = & \Omega^{-1}\left[\begin{array}{c}
\frac{\partial}{\partial\bfA_{J}}\\
\frac{\partial}{\partial\bfx_{s}}\\
\frac{\partial}{\partial\dot{\bfA}_{J}}\\
\frac{\partial}{\partial\dot{\bfx}_{s}}
\end{array}\right]H~.
\end{eqnarray}
The corresponding non-canonical Poisson bracket is 
\begin{eqnarray}
\left\{ F,G\right\} =\left[\frac{\partial F}{\partial\bfA_{J}},\frac{\partial F}{\partial\bfx_{s}},\frac{\partial F}{\partial\dot{\bfA}_{J}},\frac{\partial F}{\partial\dot{\bfx}_{s}}\right]\Omega^{-1}\left[\frac{\partial G}{\partial\bfA_{J}},\frac{\partial G}{\partial\bfx_{s}},\frac{\partial G}{\partial\dot{\bfA}_{J}},\frac{\partial G}{\partial\dot{\bfx}_{s}}\right]^{T},
\end{eqnarray}
or more specifically, 
\begin{eqnarray}
\left\{ F,G\right\}  & = & \frac{1}{\Delta V}\sum_{J}\left(\frac{\partial F}{\partial\bfA_{J}}\cdot\frac{\partial G}{\partial\dot{\bfA}_{J}}-\frac{\partial F}{\partial\dot{\bfA}_{J}}\cdot\frac{\partial G}{\partial\bfA_{J}}\right)+\sum_{s}\frac{1}{m_{s}}\left(\frac{\partial F}{\partial\bfx_{s}}\cdot\frac{\partial G}{\partial\dot{\bfx}_{s}}-\frac{\partial F}{\partial\dot{\bfx}_{s}}\cdot\frac{\partial G}{\partial\bfx_{s}}\right)+\nonumber \\
 &  & \sum_{s}\frac{q_{s}}{m_{s}\Delta V}\left(\frac{\partial G}{\partial\dot{\bfx}_{s}}\cdot\sum_{J}W_{\sigma_{1J}}\left(\bfx_{s}\right)\frac{\partial F}{\partial\dot{\bfA}_{J}}-\frac{\partial F}{\partial\dot{\bfx}_{s}}\cdot\sum_{J}W_{\sigma_{1J}}\left(\bfx_{s}\right)\frac{\partial G}{\partial\dot{\bfA}_{J}}\right)+\nonumber \\
 &  & -\sum_{s}\sum_{J}\frac{q_{s}}{m_{s}^{2}}\frac{\partial F}{\partial\dot{\bfx}_{s}}\cdot[\nabla\times W_{\sigma_{1J}}\left(\bfx_{s}\right)\bfA_{J}]\times\frac{\partial G}{\partial\dot{\bfx}_{s}}~.
\end{eqnarray}

Now, we introduce two new variables $\bfE_{J}$ and $\bfB_{K}$, which
are the discrete electric field and magnetic field, 
\begin{eqnarray}
\bfE_{J}=-\dot{\bfA}_{J}~,\\
\bfB_{K}=\sum_{J}\CURLD_{KJ}\bfA_{J}~.
\end{eqnarray}
In terms of $\bfE_{J}$ and $\bfB_{K}$, the partial derivatives with
respect to $\bfA_{J}$ and $\dot{\bfA}_{J}$ are 
\begin{eqnarray}
\frac{\partial F}{\partial\bfA_{J}} & = & \sum_{K}\frac{\partial F}{\partial\bfB_{K}}\CURLD_{KJ}~,\label{eq:20}\\
\frac{\partial F}{\partial\dot{\bfA}_{J}} & = & -\frac{\partial F}{\partial\bfE_{J}}~.
\end{eqnarray}
Note that $\CURLD_{KJ}$ in Eq.\,\eqref{eq:20} is a matrix. Using
Eq.\,\eqref{EqnD1to2FORM}, the Poisson bracket in terms of $\bfE_{J}$
and $\bfB_{K}$ can be written as 
\begin{multline}
\left\{ F,G\right\} =\frac{1}{\Delta V}\sum_{J}\left(\frac{\partial F}{\partial\bfE_{J}}\cdot\sum_{K}\frac{\partial G}{\partial\bfB_{K}}\CURLD_{KJ}-\sum_{K}\frac{\partial F}{\partial\bfB_{K}}\CURLD_{KJ}\cdot\frac{\partial G}{\partial\bfE_{J}}\right)+\\
\sum_{s}\frac{1}{m_{s}}\left(\frac{\partial F}{\partial\bfx_{s}}\cdot\frac{\partial G}{\partial\dot{\bfx}_{s}}-\frac{\partial F}{\partial\dot{\bfx}_{s}}\cdot\frac{\partial G}{\partial\bfx_{s}}\right)+\\
\sum_{s}\frac{q_{s}}{m_{s}\Delta V}\left(\frac{\partial F}{\partial\dot{\bfx}_{s}}\cdot\sum_{J}W_{\sigma_{1J}}\left(\bfx_{s}\right)\frac{\partial G}{\partial\bfE_{J}}-\frac{\partial G}{\partial\dot{\bfx}_{s}}\cdot\sum_{J}W_{\sigma_{1J}}\left(\bfx_{s}\right)\frac{\partial F}{\partial\bfE_{J}}\right)+\\
-\sum_{s}\frac{q_{s}}{m_{s}^{2}}\frac{\partial F}{\partial\dot{\bfx}_{s}}\cdot\left[\sum_{K}W_{\sigma_{2K}}\left(\bfx_{s}\right)\bfB_{K}\right]\times\frac{\partial G}{\partial\dot{\bfx}_{s}}~,\label{eq:22}
\end{multline}
and the Hamiltonian is 
\begin{eqnarray}
H=\frac{1}{2}\left(\Delta V\sum_{J}\bfE_{J}^{2}+\Delta V\sum_{K}\bfB_{K}^{2}+\sum_{s}m_{s}\dot{\bfx}_{s}^{2}\right)~.\label{EqnHamitEB}
\end{eqnarray}
The evolution equations is 
\begin{eqnarray}
\dot{F}=\{F,H\}~,
\end{eqnarray}
where 
\begin{eqnarray}
F=[\bfE_{J},\bfB_{K},\bfx_{s},\dot{\bfx}_{s}]~.
\end{eqnarray}

This is a Hamiltonian system with a non-canonical symplectic structure
or Poisson bracket. In general, symplectic integrators for non-canonical
systems are difficult to construct. However, using the splitting method
discovered by He et al. \cite{he2015hamiltonian,he2015explicit},
we have found explicit high-order symplectic algorithms for this Hamiltonian
system that preserve its non-canonical symplectic structure. We note
that splitting method had been applied to the Vlasov equation previously
without the context of symplectic structure \cite{Crouseilles07}.
We split the Hamiltonian in Eq.\,\eqref{EqnHamitEB} into five parts,
\begin{eqnarray}
H=H_{E}+H_{B}+H_{x}+H_{y}+H_{z}~,\\
H_{E}=\frac{1}{2}\Delta V\sum_{J}\bfE_{J}^{2}~,\\
H_{B}=\frac{1}{2}\Delta V\sum_{K}\bfB_{K}^{2}~,\\
H_{r}=\frac{1}{2}\sum_{s}m_{s}\dot{r}_{s}^{2},\quad\textrm{for }r\textrm{ in}~x,~y,~z~.
\end{eqnarray}
It turns out that the sub-system generated by each part can be solved
exactly, and high-order symplectic algorithms follow by combination.
The evolution equation for $H_{E}$ is $\dot{F}=\{F,H_{E}\}$, which
can be written as 
\begin{eqnarray}
\dot{\bfE}_{J} & = & 0~,\\
\dot{\bfB}_{K} & = & -\sum_{J}\CURLD_{KJ}\bfE_{J}~,\\
\dot{\bfx}_{s} & = & 0~,\\
\ddot{\bfx}_{s} & = & \frac{q_{s}}{m_{s}}\sum_{J}W_{\sigma_{1J}}\left(\bfx_{s}\right)\bfE_{J}~.
\end{eqnarray}
The exact solution $\Theta_{E}\left(\Delta t\right)$ for any time
step $\Delta t$ is 
\begin{eqnarray}
\bfE_{J}\left(t+\Delta t\right) & = & \bfE_{J}\left(t\right)~,\\
\bfB_{K}\left(t+\Delta t\right) & = & \bfB_{K}\left(t\right)-\Delta t\sum_{J}\CURLD_{KJ}\bfE_{J}(t)~,\\
\bfx_{s}\left(t+\Delta t\right) & = & \bfx_{s}\left(t\right)~,\\
\dot{\bfx}_{s}\left(t+\Delta t\right) & = & \dot{\bfx}_{s}\left(t\right)+\frac{q_{s}}{m_{s}}\Delta t\sum_{J}W_{\sigma_{1J}}\left(\bfx_{s}(t)\right)\bfE_{J}(t)~.
\end{eqnarray}
The evolution equation for $H_{B}$ is $\dot{F}=\{F,H_{B}\}$, or
\begin{eqnarray}
\dot{\bfE}_{J} & = & \sum_{K}\CURLD_{KJ}\bfB_{K}~,\\
\dot{\bfB}_{K} & = & 0~,\\
\dot{\bfx}_{s} & = & 0~,\\
\ddot{\bfx}_{s} & = & 0~,
\end{eqnarray}
whose exact solution $\Theta_{B}\left(\Delta t\right)$ is 
\begin{eqnarray}
\bfE_{J}\left(t+\Delta t\right) & = & \bfE_{J}\left(t\right)+\Delta t\sum_{K}\CURLD_{KJ}\bfB_{K}(t)~,\\
\bfB_{K}\left(t+\Delta t\right) & = & \bfB_{K}\left(t\right)~,\\
\bfx_{s}\left(t+\Delta t\right) & = & \bfx_{s}\left(t\right)~,\\
\dot{\bfx}_{s}\left(t+\Delta t\right) & = & \dot{\bfx}_{s}\left(t\right)~.
\end{eqnarray}
The evolution equation for $H_{x}$ is $\dot{F}=\{F,H_{x}\}$, or
\begin{eqnarray}
\dot{\bfE}_{J} & = & -\sum_{s}\frac{q_{s}}{\Delta V}\dot{x}_{s}\bfe_{x}W_{\sigma_{1J}}\left(\bfx_{s}\right)~,\\
\dot{\bfB}_{K} & = & 0~,\\
\dot{\bfx}_{s} & = & \dot{x}_{s}\bfe_{x}~,\\
\ddot{\bfx}_{s} & = & \frac{q_{s}}{m_{s}}\dot{x}_{s}\bfe_{x}\times\sum_{K}W_{\sigma_{2K}}\left(\bfx_{s}\right)\bfB_{K}~.
\end{eqnarray}
The exact solution $\Theta_{x}\left(\Delta t\right)$ of this sub-system
can also be computed as 
\begin{eqnarray}
\bfE_{J}\left(t+\Delta t\right) & = & \bfE_{J}\left(t\right)-\int_{0}^{\Delta t}dt'\sum_{s}\frac{q_{s}}{\Delta V}\dot{x}_{s}(t)\bfe_{x}W_{\sigma_{1J}}\left(\bfx_{s}\left(t\right)+\dot{x}_{s}(t)t'\bfe_{x}\right)~,\\
\bfB_{K}\left(t+\Delta t\right) & = & \bfB_{K}\left(t\right)~,\\
\bfx_{s}\left(t+\Delta t\right) & = & \bfx_{s}\left(t\right)+\Delta t\dot{x}_{s}(t)\bfe_{x}~,\\
\dot{\bfx}_{s}\left(t+\Delta t\right) & = & \dot{\bfx}_{s}\left(t\right)+\frac{q_{s}}{m_{s}}\dot{x}_{s}(t)\bfe_{x}\times\int_{0}^{\Delta t}dt'\sum_{K}W_{\sigma_{2K}}\left(\bfx_{s}\left(t\right)+\dot{x}_{s}(t)t'\bfe_{x}\right)\bfB_{K}(t)~.
\end{eqnarray}
Exact solutions $\Theta_{y}\left(\Delta t\right)$ and $\Theta_{z}\left(\Delta t\right)$
for sub-systems corresponding to $H_{y}$ and $H_{z}$ are obtained
in a similar manner. These exact solutions for sub-systems are then
combined to construct symplectic integrators for the original non-canonical
Hamiltonian system specified by Eqs.\,\eqref{eq:22} and \eqref{EqnHamitEB}.
For example, a first order scheme can be constructed as 
\begin{eqnarray}
\Theta_{1}\left(\Delta t\right)=\Theta_{E}\left(\Delta t\right)\Theta_{B}\left(\Delta t\right)\Theta_{x}\left(\Delta t\right)\Theta_{y}\left(\Delta t\right)\Theta_{z}\left(\Delta t\right)~,
\end{eqnarray}
and a second order symmetric scheme is

\begin{eqnarray}
\Theta_{2}\left(\Delta t\right) & = & \Theta_{x}\left(\Delta t/2\right)\Theta_{y}\left(\Delta t/2\right)\Theta_{z}\left(\Delta t/2\right)\Theta_{B}\left(\Delta t/2\right)\Theta_{E}\left(\Delta t\right)\nonumber \\
 &  & \Theta_{B}\left(\Delta t/2\right)\Theta_{z}\left(\Delta t/2\right)\Theta_{y}\left(\Delta t/2\right)\Theta_{x}\left(\Delta t/2\right)~.
\end{eqnarray}
An algorithm with order $2(l+1)$ can be constructed in the following
way,

\begin{eqnarray}
\Theta_{2(l+1)}(\Delta t) & = & \Theta_{2l}(\alpha_{l}\Delta t)\Theta_{2l}(\beta_{l}\Delta t)\Theta_{2l}(\alpha_{l}\Delta t)~,\\
\alpha_{l} & = & 1/(2-2^{1/(2l+1)})~,\\
\beta_{l} & = & 1-2\alpha_{l}~.
\end{eqnarray}

\section{Numerical Examples}

\begin{figure}
\begin{centering}
\includegraphics[width=0.5\linewidth]{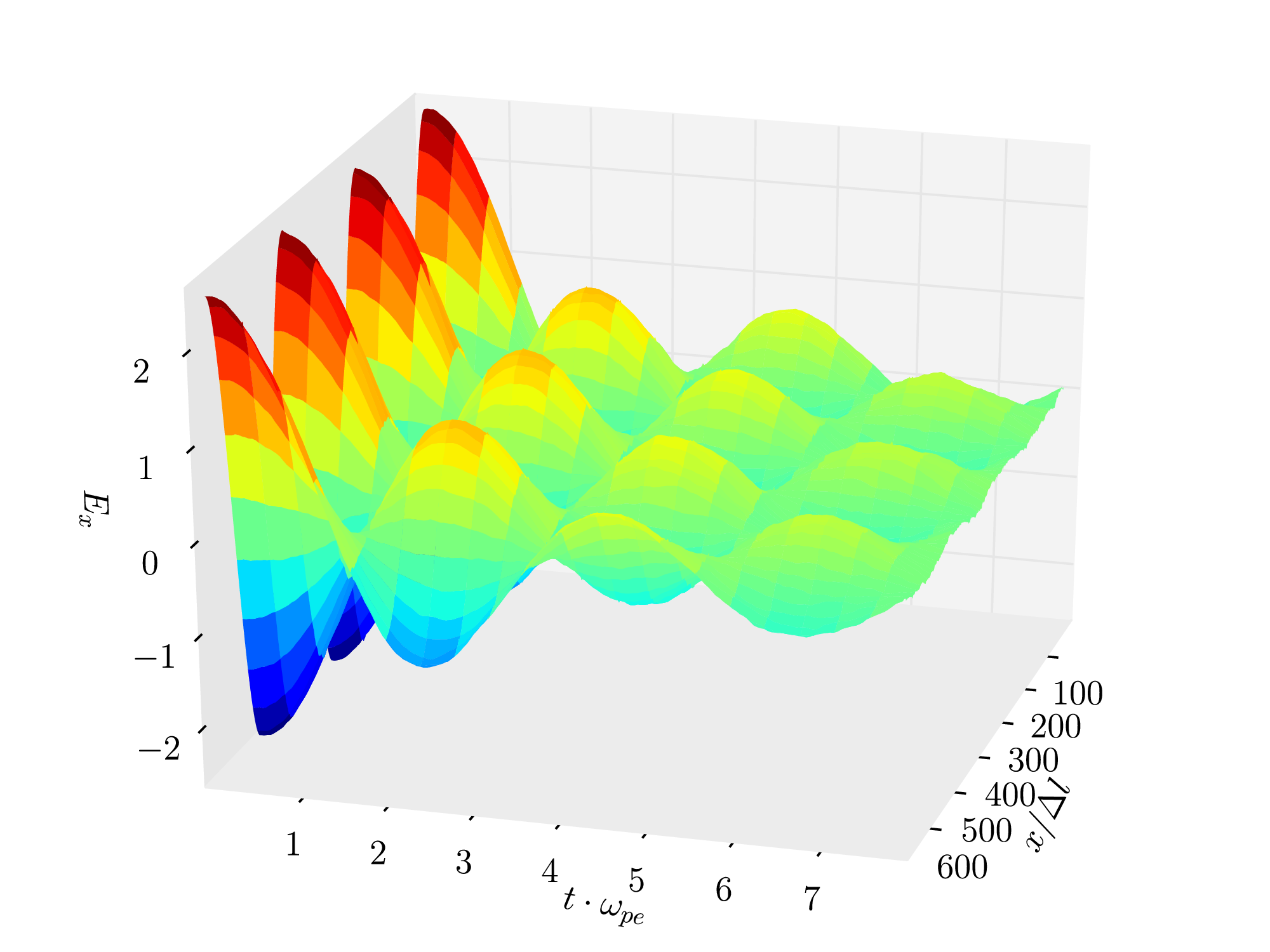} 
\par\end{centering}

\caption{The time evolution of an electrostatic wave in a hot plasma.}
\label{FigLandauEx} 
\end{figure}

\begin{figure}
\begin{centering}
\includegraphics[width=0.5\linewidth]{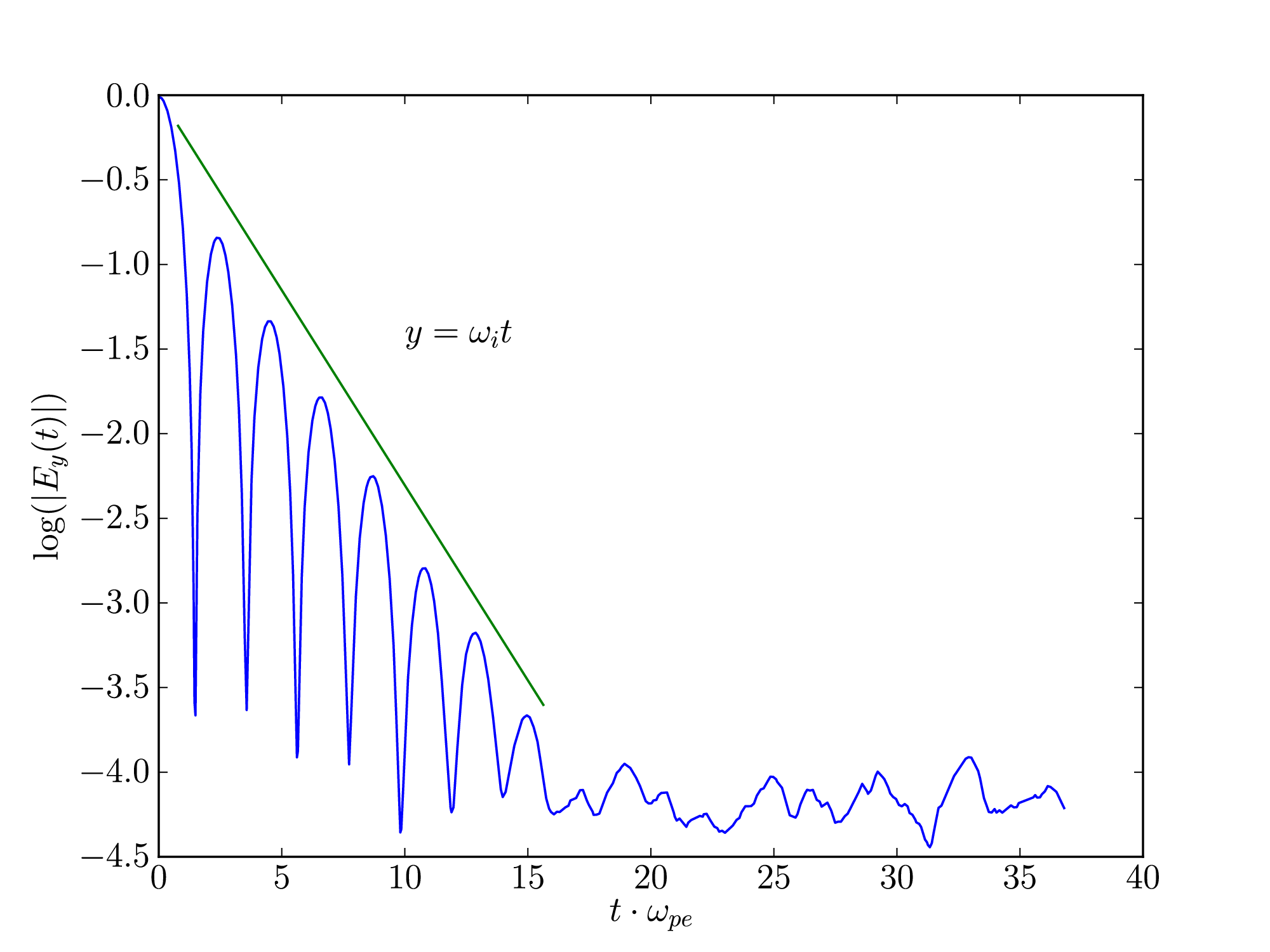} 
\par\end{centering}

\caption{Logarithmic plot of the time evolution of absolute value of the electric
field. The slope of the solid green line is the theoretical damping
rate.}
\label{FigLandamping} 
\end{figure}

We have implement the second-order non-canonical symplectic PIC algorithm
described above using the C programming language. The code is parallelized
using MPI and OpenMP. To test the algorithm, two physics problems
are simulated. The first problem is the nonlinear Landau damping of
an electrostatic wave in a hot plasma, which has been investigated
theoretically \cite{dawson1961landau,o1965collisionless,mouhot2011landau}
and numerically \cite{manfredi1997long,zhou2001numerical,kraus2013variational}.
The density of electron $n_{e}$ is $1.2116\EXP^{16}\mathrm{m}^{-3}$,
and the electron velocity is Maxwellian distributed with thermal speed
$v_{T}=0.1c$, where $c$ is the speed of light in vacuum. The computation
is carried out in a $672\times1\times1$ cubic mesh, and the size
of each grid cell is $\Delta l=2.4355\EXP^{-4}\mathrm{m}$. There
is no external electromagnetic field, and there are 40000 sample particles
in each cell when unperturbed. The initial electric field is $\bfE_{1}=E_{1}\cos\left(kx\right)\bfe_{x}$,
where $k=2\pi/224\Delta l$ is the wave number, and the amplitude
is $E_{1}=36$kV/m. The simulation is carried out for 15000 time-steps,
and the electric field is recorded during the simulation. We plot
the evolution of electric field to observe the Landau damping phenomenon
(Figs.~\ref{FigLandauEx} and \ref{FigLandamping}). The theoretical
damping rate of electric field is $\omega_{i}=-1.3223\EXP^{9}\mathrm{rad/s}$,
and it is evident that the simulation result agrees well with the
theory. 

\begin{figure}
\begin{centering}
\includegraphics[width=0.5\linewidth]{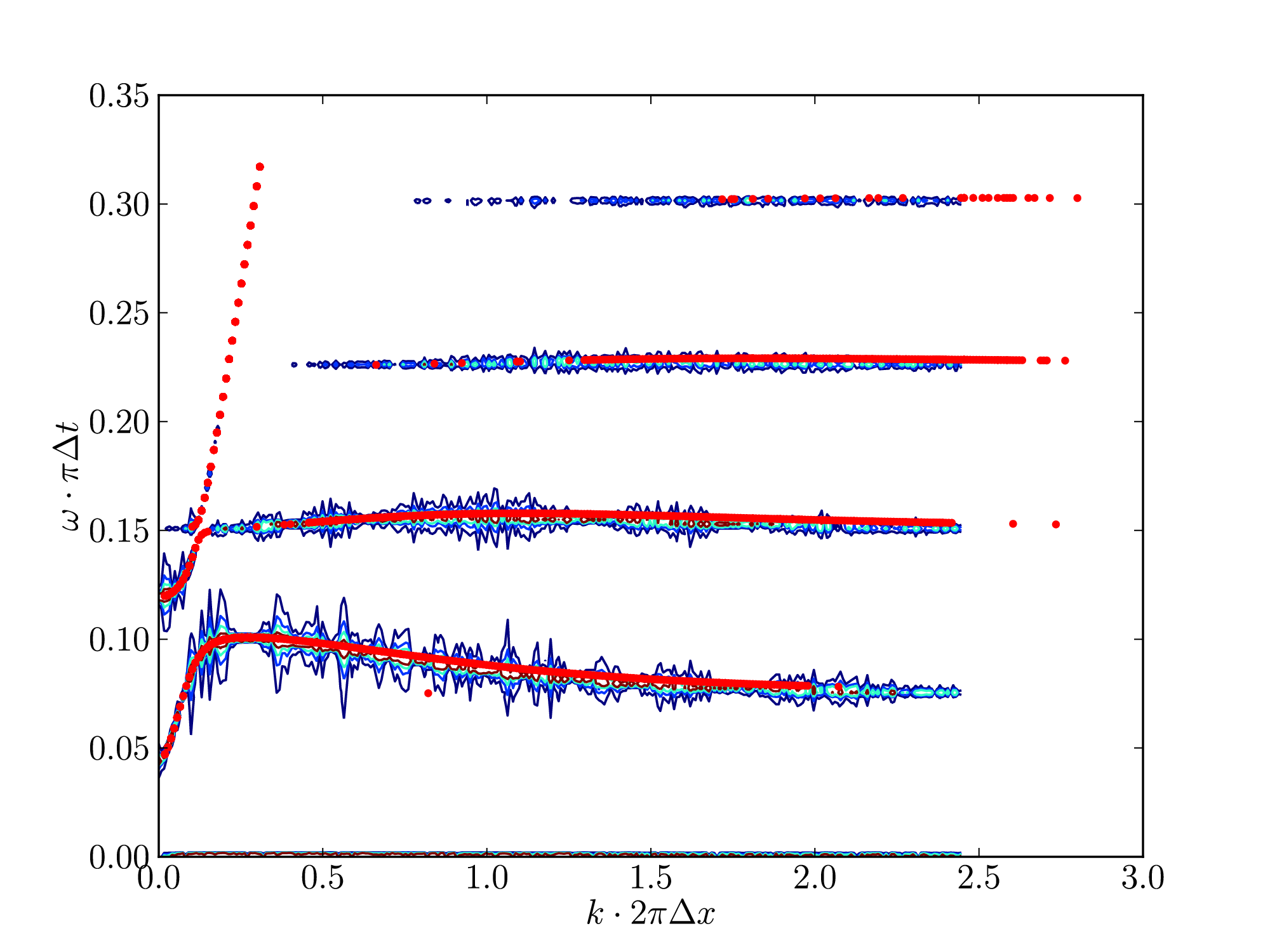} 
\par\end{centering}

\caption{The dispersion relation of electron Bernstein wave (contours) obtained
by the non-canonical symplectic PIC method. The red dots are calculated
from the analytical dispersion relation.}
\label{FigEBWDisp} 
\end{figure}

Another test problem is the dispersion relation of electron Bernstein
waves \cite{stix1992waves}. In this problem, an electromagnetic wave
propagates perpendicularly to an uniform external magnetic field $\bfB_{0}=B_{0}\bfe_{z}$
with $B_{0}=5.13$T. Other system parameters are 
\begin{eqnarray}
n_{e} & = & 2.4\EXP^{20}\mathrm{m^{-3}}~,\\
v_{T} & = & 0.07c~,\\
\Delta l & = & 2.5\EXP^{-5}m~,\\
\Delta t & = & \frac{\Delta l}{2c}~.
\end{eqnarray}
The computation domain is a $768\times1\times1$ cubic mesh, and the
averaged number of sample points per grid is 4000. An initial electromagnetic
perturbation is imposed, and after simulating 6000 time steps the
space-time spectrum of $E_{x}$ is plotted in Fig.\,\ref{FigEBWDisp},
which shows that the dispersion relation simulated matches the theoretical
curve perfectly.

\begin{figure}
\begin{centering}
\subfloat[Second order algorithm.]{\begin{centering}
\includegraphics[width=0.49\linewidth]{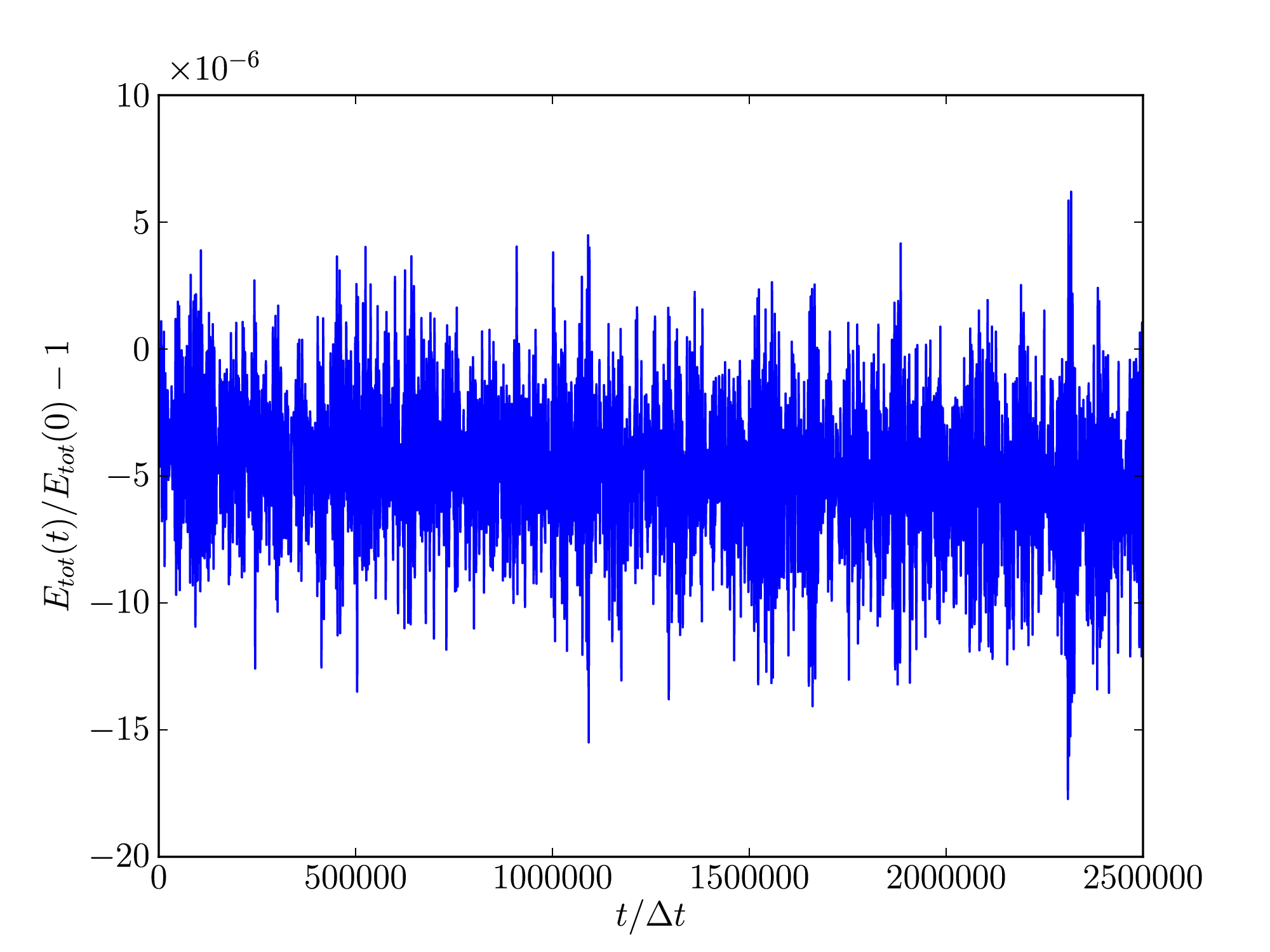}
\par\end{centering}

}\subfloat[First order algorithm.]{\begin{centering}
\includegraphics[width=0.49\linewidth]{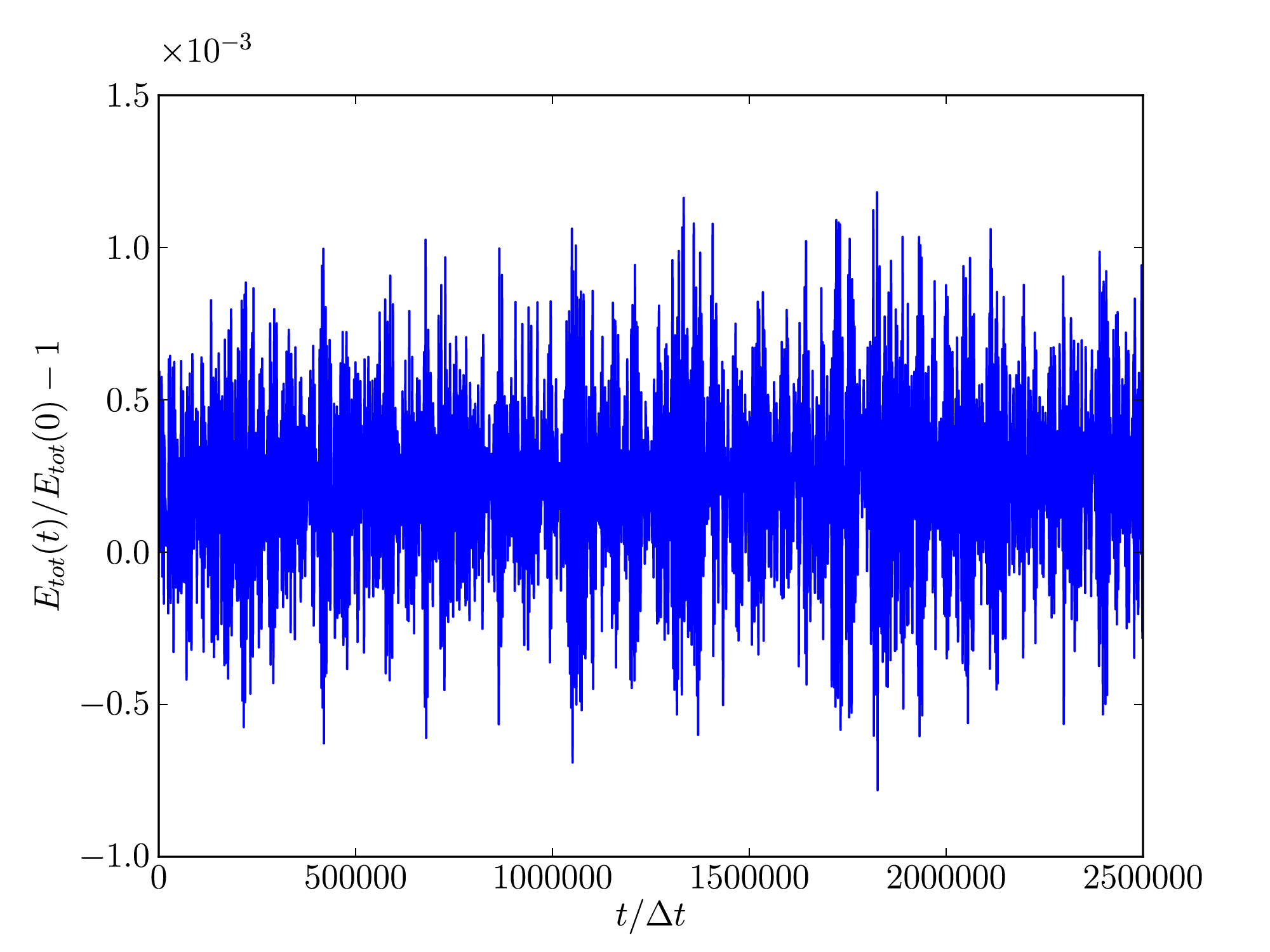}
\par\end{centering}

}
\par\end{centering}

\caption{Total energy error as a function of time for a magnetized hot plasma
obtained by the second (a) and first (b) order non-canonical symplectic
PIC method. Both energy errors are bounded within a small value. }
\label{FigEBWENE} 
\end{figure}

As a symplectic method, the non-canonical symplectic PIC algorithm
is expected to have good long-term properties. To demonstrate that,
we run another test, where the system parameters are same as the previous
problem for the Bernstein wave, except that the number of sample particles
is 40 per grid and the simulation domain is a $48\times1\times1$
cubic mesh. Both the second order and the first order split methods
are tested. The simulation is run for 2.5 million time-steps, and
the evolution of total energy errors are plotted in Fig.\,\ref{FigEBWENE}.
It is clear that the energy errors are bounded within a small value
during the entire long-term simulation for both split methods, and
the second order method is more accurate than the first order method.

\section{Summary and Discussion}

We have developed and tested a non-canonical symplectic PIC algorithm
for the VM system. The non-canonical symplectic structure is obtained
by discretizing the electromagnetic field of the particle-field Lagrangian
using the method of discrete exterior calculus. A high-order interpolating
method for differential forms is developed to render smooth interpolations
of the electromagnetic field. The effectiveness and conservative nature
of the algorithm has been verified by the physics problems of nonlinear
Landau damping and electron Bernstein wave.

\appendix

\section{High order interpolation forms for a cubic mesh}

\label{SecWhitneyInterp} The interpolation forms for a cubic mesh
is inspired by the Whitney forms \cite{whitney1957geometric}, which
were originally developed as first order interpolation forms over
simplicial complex and has become an important tool in discrete exterior
calculus (DEC) \cite{hirani2003discrete,whitney1957geometric,desbrun2008discrete}.
In DEC theory, the discrete forms are defined on chains \cite{hirani2003discrete}.
For example, discrete 0-forms are defined on vertexes of the grid,
discrete 1-forms are defined on edges, and the discrete differential
operators such as $\nabla_{d}$ and $\CURLD$ are discrete exterior
derivatives $\bfd_{d}$ acted on these discrete forms. The Whitney
map $\phi_{W}$ is a map that allows us to define continuous forms
based on these discrete forms \cite{hirani2003discrete}. With this
map, the following relation holds for any discrete form $\alpha$,
\begin{eqnarray}
\phi_{W}\bfd_{d}\alpha=\bfd\phi_{W}\alpha~,
\end{eqnarray}
where $\bfd$ is the continuous exterior derivative.

DEC solvers for Maxwell equations in cubic meshes are given by Stern,
et al. \cite{stern2007geometric}. For our purpose, we need to construct
appropriate discrete differential operators $\nabla_{\mathrm{d}}$,
$\CURLD$, $\DIVD$ as well as interpolation functions $W_{\sigma_{0I}}\left(\bfx\right)$,
$W_{\sigma_{1J}}\left(\bfx\right)$, $W_{\sigma_{2K}}\left(\bfx\right)$
and $W_{\sigma_{3L}}\left(\bfx\right)$ in a cubic mesh such that
\begin{eqnarray}
\nabla\sum_{I}W_{\sigma_{0I}}\left(\bfx\right) & \phi_{I}= & \sum_{I,J}W_{\sigma_{1J}}\left(\bfx\right){\nabla_{\mathrm{d}}}_{JI}\phi_{I}~,\label{EqnD0to1FORMAPP}\\
\nabla\times\sum_{J}W_{\sigma_{1J}}\left(\bfx\right)\bfA_{J} & = & \sum_{J,K}W_{\sigma_{2K}}\left(\bfx\right){\CURLD}_{KJ}\bfA_{J}~,\label{EqnD1to2FORMAPP}\\
\nabla\cdot\sum_{K}W_{\sigma_{2K}}\left(\bfx\right)\bfB_{K} & = & \sum_{K,L}W_{\sigma_{3L}}\left(\bfx\right){\DIVD}_{LK}\bfB_{K}~,\label{EqnD2to3FORMAPP}
\end{eqnarray}
hold for any $\phi_{I}$, $\bfA_{J}$, and $\bfB_{K}$. 

To accomplish this goal, we start from choosing an interpolation function
$W_{\sigma_{0I}}\left(\bfx\right)=W_{\sigma_{0}}\left(\bfx-\bfx_{I}\right)$
for 0-forms (e.g. scalar potential) as follows, 
\begin{eqnarray}
W_{\sigma_{0}}\left(\bfx\right)=W_{1}\left(x\right)W_{1}\left(y\right)W_{1}\left(z\right)~,
\end{eqnarray}
where $\bfx_{I}$ is the coordinate of the $I$-th grid vertex and
the cell size is chosen to be 1 for simplicity. It is required that
$W_{1}\left(x\right)$ satisfies the following conditions, 
\begin{eqnarray}
W_{1}\left(x\right)=0,\quad\textrm{if }|x|>=1~,\\
W_{1}\left(x\right)+W_{1}\left(x-1\right)=1,\quad\textrm{if }0\leq x<1~.
\end{eqnarray}
For example, $W_{1}$ can be chosen to be piece-wise linear over one
grid cell, i.e., 
\begin{eqnarray}
W_{1}\left(x\right) & = & \left\{ \begin{array}{lc}
0, & |x|\geq1~,\\
1-|x|, & |x|<1~.
\end{array}\right.
\end{eqnarray}
\global\long\def\ORDER#1{^{(#1)}}
 For $x,y,z$ in $[i',i'+1),[j',j'+1),[k',k'+1)$, the $x$ component
of the left hand side of Eq.\,\eqref{EqnD0to1FORMAPP} $T_{x}$ is
\begin{eqnarray}
T_{x} & = & \sum_{i\in\{i',i'+1\},j\in\{j',j'+1\},k\in\{k',k'+1\}}\phi_{i,j,k}W_{1}'(x-i)W_{1}(y-j)W_{1}(z-k)\nonumber \\
 & = & \sum_{j\in\{j',j'+1\},k\in\{k',k'+1\}}\left(\phi_{i',j,k}W_{1}'(x-i')+\phi_{i'+1,j,k}W_{1}'(x-i'-1)\right)W_{1}(y-j)W_{1}(z-k)\nonumber \\
 & = & \sum_{j\in\{j',j'+1\},k\in\{k',k'+1\}}\left(\phi_{i',j,k}-\phi_{i'+1,j,k}\right)W_{1}'(x-i')W_{1}(y-j)W_{1}(z-k)~.\label{eq:72}
\end{eqnarray}
The $y$ and $z$ component can be also deduced in the similar way,
\begin{eqnarray}
T_{y} & = & \sum_{i\in\{i',i'+1\},k\in\{k',k'+1\}}\left(\phi_{i,j',k}-\phi_{i,j'+1,k}\right)W_{1}(x-i)W_{1}'(y-j')W_{1}(z-k)~,\\
T_{z} & = & \sum_{i\in\{i',i'+1\},j\in\{j',j'+1\}}\left(\phi_{i,j,k'}-\phi_{i,j,k'+1}\right)W_{1}(x-i)W_{1}(y-j)W_{1}'(z-k')~.\label{eq:74}
\end{eqnarray}
Equations \eqref{eq:72}-\eqref{eq:74} indicate that $W_{1}'(x-i)W_{1}(y-j)W_{1}(z-k)dx$,
$W_{1}(x-i)W_{1}'(y-j)W_{1}(z-k)dy$, and $W_{1}(x-i)W_{1}(y-j)W_{1}'(z-k)dz$
can be viewed as the bases for 1-form interpolation map, and that
the discrete gradient operator $\nabla_{\mathrm{d}}$ can be defined
as linear operator on $\phi_{I}$ as 
\begin{eqnarray}
 &  & \left({\nabla_{\mathrm{d}}}\phi\right)_{i,j,k}=[\phi_{i+1,j,k}-\phi_{i,j,k},\phi_{i,j+1,k}-\phi_{i,j,k},\phi_{i,j,k+1}-\phi_{i,j,k}]~.\label{EqnDEFGRADD}
\end{eqnarray}
For a given discrete 1-form field $\bfA_{I}$, the interpolated 1-form
field is 
\begin{multline}
\sum_{i,j,k}\left[{A_{x}}_{i,j,k}W_{1}\ORDER 1(x-i)W_{1}(y-j)W_{1}(z-k)dx\right.\\
+{A_{y}}_{i,j,k}W_{1}(x-i)W_{1}\ORDER 1(y-j)W_{1}(z-k)dy\\
\left.+{A_{z}}_{i,j,k}W_{1}(x-i)W_{1}(y-j)W_{1}\ORDER 1(z-k)dz\right],\label{eq:75}
\end{multline}
where 
\begin{eqnarray}
W_{1}\ORDER 1\left(x\right)=\left\{ \begin{array}{lc}
-W_{1}'(x), & 0\leq x<1~,\\
0, & \textrm{otherwise}~.
\end{array}\right.\label{eq:76}
\end{eqnarray}
is the one-cell interpolation function. The components of this interpolated
1-form field are written as

\begin{equation}
\sum_{i,j,k}W_{\sigma_{1,i,j,k}}\left(\bfx\right)\bfA_{i,j,k}\equiv\sum_{i,j,k}\left[\begin{array}{c}
{A_{x}}_{i,j,k}W_{1}\ORDER 1(x-i)W_{1}(y-j)W_{1}(z-k)\\
{A_{y}}_{i,j,k}W_{1}(x-i)W_{1}\ORDER 1(y-j)W_{1}(z-k)\\
{A_{z}}_{i,j,k}W_{1}(x-i)W_{1}(y-j)W_{1}\ORDER 1(z-k)~
\end{array}\right]^{T}~.\label{eq:EqnDEFCONTGRAD}
\end{equation}

By the same procedure, we find that discrete differential operators
$\CURLD$ and $\DIVD$ should be defined as 
\begin{eqnarray}
 &  & \left({\CURLD}\bfA\right)_{i,j,k}=\left[\begin{array}{c}
\left({A_{z}}_{i,j+1,k}-{A_{z}}_{i,j,k}\right)-\left({A_{y}}_{i,j,k+1}-{A_{y}}_{i,j,k}\right)\\
\left({A_{x}}_{i,j,k+1}-{A_{x}}_{i,j,k}\right)-\left({A_{z}}_{i+1,j,k}-{A_{z}}_{i,j,k}\right)\\
\left({A_{y}}_{i+1,j,k}-{A_{y}}_{i,j,k}\right)-\left({A_{x}}_{i,j+1,k}-{A_{x}}_{i,j,k}\right)
\end{array}\right]^{T}~,\label{EqnDEFCURLD}\\
 &  & \left({\DIVD}\bfB\right)_{i,j,k}=\left[\left({B_{x}}_{i+1,j,k}-{B_{x}}_{i,j,k}\right)+\left({B_{y}}_{i,j+1,k}-{B_{y}}_{i,j,k}\right)+\left({B_{z}}_{i,j,k+1}-{B_{z}}_{i,j,k}\right)\right]~.\label{EqnDEFDIVD}
\end{eqnarray}

For a given discrete 2-form field $\mathbf{B}_{I}$, the interpolated
2-form field is 
\begin{multline}
\sum_{i,j,k}\left[{B_{x}}_{i,j,k}W_{1}(x-i)W_{1}\ORDER 1(y-j)W_{1}\ORDER 1(z-k)dy\land dz\right.\\
+{B_{y}}_{i,j,k}W_{1}\ORDER 1(x-i)W_{1}(y-j)W_{1}\ORDER 1(z-k)dz\land dx\\
\left.+{B_{z}}_{i,j,k}W_{1}\ORDER 1(x-i)W_{1}\ORDER 1(y-j)W_{1}(z-k)dx\land dy\right],\label{eq:80}
\end{multline}
whose components can be written as 
\begin{equation}
\sum_{i,j,k}W_{\sigma_{2,i,j,k}}\left(\bfx\right)\bfB_{i,j,k}\equiv\sum_{i,j,k}\left[\begin{array}{c}
{B_{x}}_{i,j,k}W_{1}(x-i)W_{1}\ORDER 1(y-j)W_{1}\ORDER 1(z-k)\\
{B_{y}}_{i,j,k}W_{1}\ORDER 1(x-i)W_{1}(y-j)W_{1}\ORDER 1(z-k)\\
{B_{z}}_{i,j,k}W_{1}\ORDER 1(x-i)W_{1}\ORDER 1(y-j)W_{1}(z-k)
\end{array}\right]^{T}.\label{EqnDEFCONTCURL}
\end{equation}

For a discrete 3-form field $\rho_{I}$, the interpolated 3-form field
is 
\begin{equation}
\sum_{i,j,k}\rho_{i,j,k}W_{1}\ORDER 1(x-i)W_{1}\ORDER 1(y-j)W_{1}\ORDER 1(z-k)dx\land dy\land dz~,\label{eq:82}
\end{equation}
and the corresponding scaler is denoted as
\begin{equation}
\sum_{i,j,k}W_{\sigma_{3,i,j,k}}\left(\bfx\right)\rho_{i,j,k}\equiv\sum_{i,j,k}\rho_{i,j,k}W_{1}\ORDER 1(x-i)W_{1}\ORDER 1(y-j)W_{1}\ORDER 1(z-k)~.\label{EqnDEFCONTDIV}
\end{equation}

We can verify that 
\begin{eqnarray}
\left(\CURLD\nabla_{\mathrm{d}}\phi\right)_{i,j,k}=0,\quad\textrm{for any }i,j,k\textrm{ and }\phi~,\label{eq:84}\\
\left(\DIVD\CURLD\bfA\right)_{i,j,k}=0,\quad\textrm{for any }i,j,k\textrm{ and }\bfA~.\label{eq:85}
\end{eqnarray}

In Ref. \cite{stern2007geometric} the discrete electromagnetic fields
in cubic mesh seem different from ours on first look. But the difference
is merely in the notation for indices. For example, for discrete 1-form,
we can alternatively use half-integer indices to rewrite Eq. (\ref{EqnDEFGRADD})
as

\begin{align}
 & [\left({\nabla_{\mathrm{d}}}\phi\right)_{xi+1/2,j,k},\left({\nabla_{\mathrm{d}}}\phi\right)_{yi,j+1/2,k},\left({\nabla_{\mathrm{\mathrm{d}}}}\phi\right)_{zi,j,k+1/2}]=\nonumber \\
 & [\phi_{i+1,j,k}-\phi_{i,j,k},\phi_{i,j+1,k}-\phi_{i,j,k},\phi_{i,j,k+1}-\phi_{i,j,k}]~,
\end{align}
which is then identical with the notation in Ref. \cite{stern2007geometric}. 

The above interpolation forms are defined over one grid cell. For
the simulations reported here, in order to achieve higher accuracy,
we have developed and deployed high-order interpolation forms over
two grid cells. The interpolation 0-forms are 
\begin{eqnarray}
W_{\sigma_{0}}\left(\bfx\right)=W_{1}\left(x\right)W_{1}\left(y\right)W_{1}\left(z\right)~,
\end{eqnarray}
where $W_{1}\left(x\right)$ satisfies 
\begin{eqnarray}
W_{1}\left(x\right) & = & 0,\quad\textrm{if }|x|>=2~,\\
W_{1}\left(x+1\right)+W_{1}\left(x\right)+W_{1}\left(x-1\right)+W_{1}\left(x-2\right) & = & 1,\quad\textrm{if }0\leq x<1~.
\end{eqnarray}
The $W_{1}$ adopted in the algorithm is 
\begin{eqnarray}
W_{1}\left(x\right)=\left\{ \begin{array}{lc}
0~, & x>2~,\\
\frac{15}{1024}x^{8}-\frac{15}{128}x^{7}+\frac{49}{128}x^{6}-\frac{21}{32}x^{5}+\frac{35}{64}x^{4}-x+1~, & 1<x\leq2~,\\
-\frac{15}{1024}x^{8}-\frac{15}{128}x^{7}+\frac{7}{16}x^{6}-\frac{21}{32}x^{5}+\frac{175}{256}x^{4}-\frac{105}{128}x^{2}+\frac{337}{512}~, & 0<x\leq1~,\\
-\frac{15}{1024}x^{8}+\frac{15}{128}x^{7}+\frac{7}{16}x^{6}+\frac{21}{32}x^{5}+\frac{175}{256}x^{4}-\frac{105}{128}x^{2}+\frac{337}{512}~, & -1<x\leq0~,\\
\frac{15}{1024}x^{8}+\frac{15}{128}x^{7}+\frac{49}{128}x^{6}+\frac{21}{32}x^{5}+\frac{35}{64}x^{4}+x+1~, & -2<x\leq-1~,\\
0~, & x\leq-2~.
\end{array}\right.
\end{eqnarray}
It can be proved that this piece-wise polynomial function is 3rd order
continuous in the whole space. For this two-cell interpolation scheme,
the $\nabla_{\mathrm{d}}$ defined in Eq.\,(\ref{EqnDEFGRADD}),
$\CURLD$ defined in Eq.\,(\ref{EqnDEFCURLD}), and $\DIVD$ defined
in Eq.\,(\ref{EqnDEFDIVD}) remain the same, but the function $W_{1}\ORDER 1$
in Eqs.\,\eqref{eq:75}, \eqref{eq:EqnDEFCONTGRAD}, \eqref{eq:80},
(\ref{EqnDEFCONTCURL}), \eqref{eq:82} and (\ref{EqnDEFCONTDIV})
needs to be replaced by the $W_{1}^{(2)}$ function defined as 
\begin{eqnarray}
W_{1}\ORDER 2\left(x\right)=-\left\{ \begin{array}{lc}
W_{1}'\left(x\right)+W_{1}'\left(x+1\right)+W_{1}'\left(x+2\right)~, & -1\leq x<2~,\\
0~, & \mathrm{otherwise}~.
\end{array}\right.
\end{eqnarray}
It can be proved that Eqs.\,(\ref{EqnD0to1FORMAPP}), (\ref{EqnD1to2FORMAPP}),
(\ref{EqnD2to3FORMAPP}), \eqref{eq:84}, and \eqref{eq:85} hold
for this two-cell interpolation scheme.
\begin{acknowledgments}
This research is supported by ITER-China Program (2015GB111003, 2014GB124005,
2013GB111000), JSPS-NRF-NSFC A3 Foresight Program in the field of
Plasma Physics (NSFC-11261140328), the National Science Foundation
of China (11575186, 11575185, 11505185, 11505186), the CAS Program
for Interdisciplinary Collaboration Team, the Geo-Algorithmic Plasma
Simulator (GAPS) project, and the U.S. Department of Energy (DE-AC02-09CH11466).
\end{acknowledgments}

\bibliographystyle{apsrev4-1}
\bibliography{sympic2.bib}

\begin{thebibliography}{57}%
\makeatletter
\providecommand \@ifxundefined [1]{%
 \@ifx{#1\undefined}
}%
\providecommand \@ifnum [1]{%
 \ifnum #1\expandafter \@firstoftwo
 \else \expandafter \@secondoftwo
 \fi
}%
\providecommand \@ifx [1]{%
 \ifx #1\expandafter \@firstoftwo
 \else \expandafter \@secondoftwo
 \fi
}%
\providecommand \natexlab [1]{#1}%
\providecommand \enquote  [1]{``#1''}%
\providecommand \bibnamefont  [1]{#1}%
\providecommand \bibfnamefont [1]{#1}%
\providecommand \citenamefont [1]{#1}%
\providecommand \href@noop [0]{\@secondoftwo}%
\providecommand \href [0]{\begingroup \@sanitize@url \@href}%
\providecommand \@href[1]{\@@startlink{#1}\@@href}%
\providecommand \@@href[1]{\endgroup#1\@@endlink}%
\providecommand \@sanitize@url [0]{\catcode `\\12\catcode `\$12\catcode
  `\&12\catcode `\#12\catcode `\^12\catcode `\_12\catcode `\%12\relax}%
\providecommand \@@startlink[1]{}%
\providecommand \@@endlink[0]{}%
\providecommand \url  [0]{\begingroup\@sanitize@url \@url }%
\providecommand \@url [1]{\endgroup\@href {#1}{\urlprefix }}%
\providecommand \urlprefix  [0]{URL }%
\providecommand \Eprint [0]{\href }%
\providecommand \doibase [0]{http://dx.doi.org/}%
\providecommand \selectlanguage [0]{\@gobble}%
\providecommand \bibinfo  [0]{\@secondoftwo}%
\providecommand \bibfield  [0]{\@secondoftwo}%
\providecommand \translation [1]{[#1]}%
\providecommand \BibitemOpen [0]{}%
\providecommand \bibitemStop [0]{}%
\providecommand \bibitemNoStop [0]{.\EOS\space}%
\providecommand \EOS [0]{\spacefactor3000\relax}%
\providecommand \BibitemShut  [1]{\csname bibitem#1\endcsname}%
\let\auto@bib@innerbib\@empty
\bibitem [{\citenamefont {Ruth}(1983)}]{Ruth83}%
  \BibitemOpen
  \bibfield  {author} {\bibinfo {author} {\bibfnamefont {R.~D.}\ \bibnamefont
  {Ruth}},\ }\href@noop {} {\bibfield  {journal} {\bibinfo  {journal} {IEEE
  Trans. Nucl. Sci}\ }\textbf {\bibinfo {volume} {30}},\ \bibinfo {pages}
  {2669} (\bibinfo {year} {1983})}\BibitemShut {NoStop}%
\bibitem [{\citenamefont {Feng}(1985)}]{Feng85}%
  \BibitemOpen
  \bibfield  {author} {\bibinfo {author} {\bibfnamefont {K.}~\bibnamefont
  {Feng}},\ }in\ \href@noop {} {\emph {\bibinfo {booktitle} {the Proceedings of
  1984 Beijing Symposium on Differential Geometry and Differential
  Equations}}},\ \bibinfo {editor} {edited by\ \bibinfo {editor} {\bibfnamefont
  {K.}~\bibnamefont {Feng}}}\ (\bibinfo  {publisher} {Science Press},\ \bibinfo
  {year} {1985})\ pp.\ \bibinfo {pages} {42--58}\BibitemShut {NoStop}%
\bibitem [{\citenamefont {Feng}(1986)}]{Feng86}%
  \BibitemOpen
  \bibfield  {author} {\bibinfo {author} {\bibfnamefont {K.}~\bibnamefont
  {Feng}},\ }\href@noop {} {\bibfield  {journal} {\bibinfo  {journal} {J.
  Comput. Maths.}\ }\textbf {\bibinfo {volume} {4}},\ \bibinfo {pages} {279}
  (\bibinfo {year} {1986})}\BibitemShut {NoStop}%
\bibitem [{\citenamefont {Feng}\ and\ \citenamefont {Qin}(2010)}]{Feng10}%
  \BibitemOpen
  \bibfield  {author} {\bibinfo {author} {\bibfnamefont {K.}~\bibnamefont
  {Feng}}\ and\ \bibinfo {author} {\bibfnamefont {M.}~\bibnamefont {Qin}},\
  }\href@noop {} {\emph {\bibinfo {title} {Symplectic Geometric Algorithms for
  Hamiltonian Systems}}}\ (\bibinfo  {publisher} {Springer-Verlag},\ \bibinfo
  {year} {2010})\BibitemShut {NoStop}%
\bibitem [{\citenamefont {Forest}\ and\ \citenamefont {Ruth}(1990)}]{Forest90}%
  \BibitemOpen
  \bibfield  {author} {\bibinfo {author} {\bibfnamefont {E.}~\bibnamefont
  {Forest}}\ and\ \bibinfo {author} {\bibfnamefont {R.~D.}\ \bibnamefont
  {Ruth}},\ }\href@noop {} {\bibfield  {journal} {\bibinfo  {journal} {Physica
  D}\ }\textbf {\bibinfo {volume} {43}},\ \bibinfo {pages} {105} (\bibinfo
  {year} {1990})}\BibitemShut {NoStop}%
\bibitem [{\citenamefont {Channell}\ and\ \citenamefont
  {Scovel}(1990)}]{Channell90}%
  \BibitemOpen
  \bibfield  {author} {\bibinfo {author} {\bibfnamefont {P.~J.}\ \bibnamefont
  {Channell}}\ and\ \bibinfo {author} {\bibfnamefont {C.}~\bibnamefont
  {Scovel}},\ }\href@noop {} {\bibfield  {journal} {\bibinfo  {journal}
  {Nonlinearity}\ }\textbf {\bibinfo {volume} {3}},\ \bibinfo {pages} {231}
  (\bibinfo {year} {1990})}\BibitemShut {NoStop}%
\bibitem [{\citenamefont {Candy}\ and\ \citenamefont {Rozmus}(1991)}]{Candy91}%
  \BibitemOpen
  \bibfield  {author} {\bibinfo {author} {\bibfnamefont {J.}~\bibnamefont
  {Candy}}\ and\ \bibinfo {author} {\bibfnamefont {W.}~\bibnamefont {Rozmus}},\
  }\href@noop {} {\bibfield  {journal} {\bibinfo  {journal} {Journal of
  Computational Physics}\ }\textbf {\bibinfo {volume} {92}},\ \bibinfo {pages}
  {230} (\bibinfo {year} {1991})}\BibitemShut {NoStop}%
\bibitem [{\citenamefont {Marsden}\ and\ \citenamefont
  {West}(2001)}]{marsden2001discrete}%
  \BibitemOpen
  \bibfield  {author} {\bibinfo {author} {\bibfnamefont {J.~E.}\ \bibnamefont
  {Marsden}}\ and\ \bibinfo {author} {\bibfnamefont {M.}~\bibnamefont {West}},\
  }\href@noop {} {\bibfield  {journal} {\bibinfo  {journal} {Acta Numer.}\
  }\textbf {\bibinfo {volume} {10}},\ \bibinfo {pages} {357} (\bibinfo {year}
  {2001})}\BibitemShut {NoStop}%
\bibitem [{\citenamefont {Hairer}\ \emph {et~al.}(2002)\citenamefont {Hairer},
  \citenamefont {Lubich},\ and\ \citenamefont {Wanner}}]{Hairer02}%
  \BibitemOpen
  \bibfield  {author} {\bibinfo {author} {\bibfnamefont {E.}~\bibnamefont
  {Hairer}}, \bibinfo {author} {\bibfnamefont {C.}~\bibnamefont {Lubich}}, \
  and\ \bibinfo {author} {\bibfnamefont {G.}~\bibnamefont {Wanner}},\
  }\href@noop {} {\emph {\bibinfo {title} {Geometric Numerical Integration:
  Structure-Preserving Algorithms for Ordinary Differential Equations}}}\
  (\bibinfo  {publisher} {Springer},\ \bibinfo {address} {New York},\ \bibinfo
  {year} {2002})\BibitemShut {NoStop}%
\bibitem [{\citenamefont {Chin}(2008)}]{chin2008symplectic}%
  \BibitemOpen
  \bibfield  {author} {\bibinfo {author} {\bibfnamefont {S.~A.}\ \bibnamefont
  {Chin}},\ }\href@noop {} {\bibfield  {journal} {\bibinfo  {journal} {Physical
  Review E}\ }\textbf {\bibinfo {volume} {77}},\ \bibinfo {pages} {066401}
  (\bibinfo {year} {2008})}\BibitemShut {NoStop}%
\bibitem [{\citenamefont {He}\ \emph {et~al.}(2015{\natexlab{a}})\citenamefont
  {He}, \citenamefont {Qin}, \citenamefont {Sun}, \citenamefont {Xiao},
  \citenamefont {Zhang},\ and\ \citenamefont {Liu}}]{he2015hamiltonian}%
  \BibitemOpen
  \bibfield  {author} {\bibinfo {author} {\bibfnamefont {Y.}~\bibnamefont
  {He}}, \bibinfo {author} {\bibfnamefont {H.}~\bibnamefont {Qin}}, \bibinfo
  {author} {\bibfnamefont {Y.}~\bibnamefont {Sun}}, \bibinfo {author}
  {\bibfnamefont {J.}~\bibnamefont {Xiao}}, \bibinfo {author} {\bibfnamefont
  {R.}~\bibnamefont {Zhang}}, \ and\ \bibinfo {author} {\bibfnamefont
  {J.}~\bibnamefont {Liu}},\ }\href@noop {} {\bibfield  {journal} {\bibinfo
  {journal} {arXiv preprint arXiv:1505.06076}\ } (\bibinfo {year}
  {2015}{\natexlab{a}})}\BibitemShut {NoStop}%
\bibitem [{\citenamefont {He}\ \emph {et~al.}(2015{\natexlab{b}})\citenamefont
  {He}, \citenamefont {Sun}, \citenamefont {Zhou}, \citenamefont {Liu},\ and\
  \citenamefont {Qin}}]{he2015explicit}%
  \BibitemOpen
  \bibfield  {author} {\bibinfo {author} {\bibfnamefont {Y.}~\bibnamefont
  {He}}, \bibinfo {author} {\bibfnamefont {Y.}~\bibnamefont {Sun}}, \bibinfo
  {author} {\bibfnamefont {Z.}~\bibnamefont {Zhou}}, \bibinfo {author}
  {\bibfnamefont {J.}~\bibnamefont {Liu}}, \ and\ \bibinfo {author}
  {\bibfnamefont {H.}~\bibnamefont {Qin}},\ }\href@noop {} {\bibfield
  {journal} {\bibinfo  {journal} {arXiv preprint arXiv:1509.07794}\ } (\bibinfo
  {year} {2015}{\natexlab{b}})}\BibitemShut {NoStop}%
\bibitem [{\citenamefont {Qin}\ and\ \citenamefont
  {Guan}(2008)}]{PhysRevLett.100.035006}%
  \BibitemOpen
  \bibfield  {author} {\bibinfo {author} {\bibfnamefont {H.}~\bibnamefont
  {Qin}}\ and\ \bibinfo {author} {\bibfnamefont {X.}~\bibnamefont {Guan}},\
  }\href {\doibase 10.1103/PhysRevLett.100.035006} {\bibfield  {journal}
  {\bibinfo  {journal} {Phys. Rev. Lett.}\ }\textbf {\bibinfo {volume} {100}},\
  \bibinfo {pages} {035006} (\bibinfo {year} {2008})}\BibitemShut {NoStop}%
\bibitem [{\citenamefont {Qin}\ \emph {et~al.}(2009)\citenamefont {Qin},
  \citenamefont {Guan},\ and\ \citenamefont {Tang}}]{qin2009variational}%
  \BibitemOpen
  \bibfield  {author} {\bibinfo {author} {\bibfnamefont {H.}~\bibnamefont
  {Qin}}, \bibinfo {author} {\bibfnamefont {X.}~\bibnamefont {Guan}}, \ and\
  \bibinfo {author} {\bibfnamefont {W.~M.}\ \bibnamefont {Tang}},\ }\href@noop
  {} {\bibfield  {journal} {\bibinfo  {journal} {Physics of Plasmas
  (1994-present)}\ }\textbf {\bibinfo {volume} {16}},\ \bibinfo {pages}
  {042510} (\bibinfo {year} {2009})}\BibitemShut {NoStop}%
\bibitem [{\citenamefont {Li}\ \emph {et~al.}(2011)\citenamefont {Li},
  \citenamefont {Qin}, \citenamefont {Pu}, \citenamefont {Xie},\ and\
  \citenamefont {Fu}}]{li2011variational}%
  \BibitemOpen
  \bibfield  {author} {\bibinfo {author} {\bibfnamefont {J.}~\bibnamefont
  {Li}}, \bibinfo {author} {\bibfnamefont {H.}~\bibnamefont {Qin}}, \bibinfo
  {author} {\bibfnamefont {Z.}~\bibnamefont {Pu}}, \bibinfo {author}
  {\bibfnamefont {L.}~\bibnamefont {Xie}}, \ and\ \bibinfo {author}
  {\bibfnamefont {S.}~\bibnamefont {Fu}},\ }\href@noop {} {\bibfield  {journal}
  {\bibinfo  {journal} {Physics of Plasmas (1994-present)}\ }\textbf {\bibinfo
  {volume} {18}},\ \bibinfo {pages} {052902} (\bibinfo {year}
  {2011})}\BibitemShut {NoStop}%
\bibitem [{\citenamefont {Squire}\ \emph
  {et~al.}(2012{\natexlab{a}})\citenamefont {Squire}, \citenamefont {Qin},\
  and\ \citenamefont {Tang}}]{squire2012geometric}%
  \BibitemOpen
  \bibfield  {author} {\bibinfo {author} {\bibfnamefont {J.}~\bibnamefont
  {Squire}}, \bibinfo {author} {\bibfnamefont {H.}~\bibnamefont {Qin}}, \ and\
  \bibinfo {author} {\bibfnamefont {W.~M.}\ \bibnamefont {Tang}},\ }\href@noop
  {} {\bibfield  {journal} {\bibinfo  {journal} {Physics of Plasmas
  (1994-present)}\ }\textbf {\bibinfo {volume} {19}},\ \bibinfo {pages}
  {084501} (\bibinfo {year} {2012}{\natexlab{a}})}\BibitemShut {NoStop}%
\bibitem [{\citenamefont {Kraus}(2013)}]{kraus2013variational}%
  \BibitemOpen
  \bibfield  {author} {\bibinfo {author} {\bibfnamefont {M.}~\bibnamefont
  {Kraus}},\ }\href@noop {} {\bibfield  {journal} {\bibinfo  {journal} {arXiv
  preprint arXiv:1307.5665}\ } (\bibinfo {year} {2013})}\BibitemShut {NoStop}%
\bibitem [{\citenamefont {Zhang}\ \emph
  {et~al.}(2014{\natexlab{a}})\citenamefont {Zhang}, \citenamefont {Liu},
  \citenamefont {Tang}, \citenamefont {Qin}, \citenamefont {Xiao},\ and\
  \citenamefont {Zhu}}]{zhang2014canonicalization}%
  \BibitemOpen
  \bibfield  {author} {\bibinfo {author} {\bibfnamefont {R.}~\bibnamefont
  {Zhang}}, \bibinfo {author} {\bibfnamefont {J.}~\bibnamefont {Liu}}, \bibinfo
  {author} {\bibfnamefont {Y.}~\bibnamefont {Tang}}, \bibinfo {author}
  {\bibfnamefont {H.}~\bibnamefont {Qin}}, \bibinfo {author} {\bibfnamefont
  {J.}~\bibnamefont {Xiao}}, \ and\ \bibinfo {author} {\bibfnamefont
  {B.}~\bibnamefont {Zhu}},\ }\href@noop {} {\bibfield  {journal} {\bibinfo
  {journal} {Physics of Plasmas (1994-present)}\ }\textbf {\bibinfo {volume}
  {21}},\ \bibinfo {pages} {032504} (\bibinfo {year}
  {2014}{\natexlab{a}})}\BibitemShut {NoStop}%
\bibitem [{\citenamefont {Ellison}\ \emph
  {et~al.}(2015{\natexlab{a}})\citenamefont {Ellison}, \citenamefont {Finn},
  \citenamefont {Qin},\ and\ \citenamefont {Tang}}]{ellison2015development}%
  \BibitemOpen
  \bibfield  {author} {\bibinfo {author} {\bibfnamefont {C.~L.}\ \bibnamefont
  {Ellison}}, \bibinfo {author} {\bibfnamefont {J.}~\bibnamefont {Finn}},
  \bibinfo {author} {\bibfnamefont {H.}~\bibnamefont {Qin}}, \ and\ \bibinfo
  {author} {\bibfnamefont {W.~M.}\ \bibnamefont {Tang}},\ }\href@noop {}
  {\bibfield  {journal} {\bibinfo  {journal} {Plasma Physics and Controlled
  Fusion}\ }\textbf {\bibinfo {volume} {57}},\ \bibinfo {pages} {054007}
  (\bibinfo {year} {2015}{\natexlab{a}})}\BibitemShut {NoStop}%
\bibitem [{\citenamefont {Qin}\ \emph {et~al.}(2013)\citenamefont {Qin},
  \citenamefont {Zhang}, \citenamefont {Xiao}, \citenamefont {Liu},
  \citenamefont {Sun},\ and\ \citenamefont {Tang}}]{qin2013boris}%
  \BibitemOpen
  \bibfield  {author} {\bibinfo {author} {\bibfnamefont {H.}~\bibnamefont
  {Qin}}, \bibinfo {author} {\bibfnamefont {S.}~\bibnamefont {Zhang}}, \bibinfo
  {author} {\bibfnamefont {J.}~\bibnamefont {Xiao}}, \bibinfo {author}
  {\bibfnamefont {J.}~\bibnamefont {Liu}}, \bibinfo {author} {\bibfnamefont
  {Y.}~\bibnamefont {Sun}}, \ and\ \bibinfo {author} {\bibfnamefont {W.~M.}\
  \bibnamefont {Tang}},\ }\href@noop {} {\bibfield  {journal} {\bibinfo
  {journal} {Physics of Plasmas (1994-present)}\ }\textbf {\bibinfo {volume}
  {20}},\ \bibinfo {pages} {084503} (\bibinfo {year} {2013})}\BibitemShut
  {NoStop}%
\bibitem [{\citenamefont {Zhang}\ \emph
  {et~al.}(2014{\natexlab{b}})\citenamefont {Zhang}, \citenamefont {Jia},\ and\
  \citenamefont {Sun}}]{zhang2015comment}%
  \BibitemOpen
  \bibfield  {author} {\bibinfo {author} {\bibfnamefont {S.}~\bibnamefont
  {Zhang}}, \bibinfo {author} {\bibfnamefont {Y.}~\bibnamefont {Jia}}, \ and\
  \bibinfo {author} {\bibfnamefont {Q.}~\bibnamefont {Sun}},\ }\href@noop {}
  {\bibfield  {journal} {\bibinfo  {journal} {Journal of Computational
  Physics}\ }\textbf {\bibinfo {volume} {282}},\ \bibinfo {pages} {43}
  (\bibinfo {year} {2014}{\natexlab{b}})}\BibitemShut {NoStop}%
\bibitem [{\citenamefont {Ellison}\ \emph
  {et~al.}(2015{\natexlab{b}})\citenamefont {Ellison}, \citenamefont {Burby},\
  and\ \citenamefont {Qin}}]{ellison2015comment}%
  \BibitemOpen
  \bibfield  {author} {\bibinfo {author} {\bibfnamefont {C.}~\bibnamefont
  {Ellison}}, \bibinfo {author} {\bibfnamefont {J.}~\bibnamefont {Burby}}, \
  and\ \bibinfo {author} {\bibfnamefont {H.}~\bibnamefont {Qin}},\ }\href
  {\doibase http://dx.doi.org/10.1016/j.jcp.2015.09.007} {\bibfield  {journal}
  {\bibinfo  {journal} {Journal of Computational Physics}\ }\textbf {\bibinfo
  {volume} {301}},\ \bibinfo {pages} {489 } (\bibinfo {year}
  {2015}{\natexlab{b}})}\BibitemShut {NoStop}%
\bibitem [{\citenamefont {He}\ \emph {et~al.}(2015{\natexlab{c}})\citenamefont
  {He}, \citenamefont {Sun}, \citenamefont {Liu},\ and\ \citenamefont
  {Qin}}]{he2015volume}%
  \BibitemOpen
  \bibfield  {author} {\bibinfo {author} {\bibfnamefont {Y.}~\bibnamefont
  {He}}, \bibinfo {author} {\bibfnamefont {Y.}~\bibnamefont {Sun}}, \bibinfo
  {author} {\bibfnamefont {J.}~\bibnamefont {Liu}}, \ and\ \bibinfo {author}
  {\bibfnamefont {H.}~\bibnamefont {Qin}},\ }\href@noop {} {\bibfield
  {journal} {\bibinfo  {journal} {Journal of Computational Physics}\ }\textbf
  {\bibinfo {volume} {281}},\ \bibinfo {pages} {135} (\bibinfo {year}
  {2015}{\natexlab{c}})}\BibitemShut {NoStop}%
\bibitem [{\citenamefont {Zhang}\ \emph {et~al.}(2015)\citenamefont {Zhang},
  \citenamefont {Liu}, \citenamefont {Qin}, \citenamefont {Wang}, \citenamefont
  {He},\ and\ \citenamefont {Sun}}]{zhang2015volume}%
  \BibitemOpen
  \bibfield  {author} {\bibinfo {author} {\bibfnamefont {R.}~\bibnamefont
  {Zhang}}, \bibinfo {author} {\bibfnamefont {J.}~\bibnamefont {Liu}}, \bibinfo
  {author} {\bibfnamefont {H.}~\bibnamefont {Qin}}, \bibinfo {author}
  {\bibfnamefont {Y.}~\bibnamefont {Wang}}, \bibinfo {author} {\bibfnamefont
  {Y.}~\bibnamefont {He}}, \ and\ \bibinfo {author} {\bibfnamefont
  {Y.}~\bibnamefont {Sun}},\ }\href@noop {} {\bibfield  {journal} {\bibinfo
  {journal} {Physics of Plasmas (1994-present)}\ }\textbf {\bibinfo {volume}
  {22}},\ \bibinfo {pages} {044501} (\bibinfo {year} {2015})}\BibitemShut
  {NoStop}%
\bibitem [{\citenamefont {Qin}\ \emph {et~al.}(2007)\citenamefont {Qin},
  \citenamefont {Cohen}, \citenamefont {Nevins},\ and\ \citenamefont
  {Xu}}]{qin2007geometric}%
  \BibitemOpen
  \bibfield  {author} {\bibinfo {author} {\bibfnamefont {H.}~\bibnamefont
  {Qin}}, \bibinfo {author} {\bibfnamefont {R.}~\bibnamefont {Cohen}}, \bibinfo
  {author} {\bibfnamefont {W.}~\bibnamefont {Nevins}}, \ and\ \bibinfo {author}
  {\bibfnamefont {X.}~\bibnamefont {Xu}},\ }\href@noop {} {\bibfield  {journal}
  {\bibinfo  {journal} {Physics of Plasmas (1994-present)}\ }\textbf {\bibinfo
  {volume} {14}},\ \bibinfo {pages} {056110} (\bibinfo {year}
  {2007})}\BibitemShut {NoStop}%
\bibitem [{\citenamefont {Qin}\ \emph {et~al.}(2014)\citenamefont {Qin},
  \citenamefont {Burby},\ and\ \citenamefont {Davidson}}]{qin2014field}%
  \BibitemOpen
  \bibfield  {author} {\bibinfo {author} {\bibfnamefont {H.}~\bibnamefont
  {Qin}}, \bibinfo {author} {\bibfnamefont {J.~W.}\ \bibnamefont {Burby}}, \
  and\ \bibinfo {author} {\bibfnamefont {R.~C.}\ \bibnamefont {Davidson}},\
  }\href@noop {} {\bibfield  {journal} {\bibinfo  {journal} {Physical Review
  E}\ }\textbf {\bibinfo {volume} {90}},\ \bibinfo {pages} {043102} (\bibinfo
  {year} {2014})}\BibitemShut {NoStop}%
\bibitem [{\citenamefont {Squire}\ \emph
  {et~al.}(2012{\natexlab{b}})\citenamefont {Squire}, \citenamefont {Qin},\
  and\ \citenamefont {Tang}}]{Squire4748}%
  \BibitemOpen
  \bibfield  {author} {\bibinfo {author} {\bibfnamefont {J.}~\bibnamefont
  {Squire}}, \bibinfo {author} {\bibfnamefont {H.}~\bibnamefont {Qin}}, \ and\
  \bibinfo {author} {\bibfnamefont {W.~M.}\ \bibnamefont {Tang}},\ }\href@noop
  {} {\emph {\bibinfo {title} {Geometric Integration Of The Vlasov-Maxwell
  System With A Variational Particle-in-cell Scheme}}},\ \bibinfo {type} {Tech.
  Rep.}\ \bibinfo {number} {PPPL-4748}\ (\bibinfo  {institution} {Princeton
  Plasma Physics Laboratory},\ \bibinfo {year} {2012})\BibitemShut {NoStop}%
\bibitem [{\citenamefont {Squire}\ \emph
  {et~al.}(2012{\natexlab{c}})\citenamefont {Squire}, \citenamefont {Qin},\
  and\ \citenamefont {Tang}}]{squire2012gauge}%
  \BibitemOpen
  \bibfield  {author} {\bibinfo {author} {\bibfnamefont {J.}~\bibnamefont
  {Squire}}, \bibinfo {author} {\bibfnamefont {H.}~\bibnamefont {Qin}}, \ and\
  \bibinfo {author} {\bibfnamefont {W.~M.}\ \bibnamefont {Tang}},\ }\href@noop
  {} {\bibfield  {journal} {\bibinfo  {journal} {Physics of Plasmas
  (1994-present)}\ }\textbf {\bibinfo {volume} {19}},\ \bibinfo {pages}
  {052501} (\bibinfo {year} {2012}{\natexlab{c}})}\BibitemShut {NoStop}%
\bibitem [{\citenamefont {Xiao}\ \emph {et~al.}(2013)\citenamefont {Xiao},
  \citenamefont {Liu}, \citenamefont {Qin},\ and\ \citenamefont
  {Yu}}]{xiao2013variational}%
  \BibitemOpen
  \bibfield  {author} {\bibinfo {author} {\bibfnamefont {J.}~\bibnamefont
  {Xiao}}, \bibinfo {author} {\bibfnamefont {J.}~\bibnamefont {Liu}}, \bibinfo
  {author} {\bibfnamefont {H.}~\bibnamefont {Qin}}, \ and\ \bibinfo {author}
  {\bibfnamefont {Z.}~\bibnamefont {Yu}},\ }\href@noop {} {\bibfield  {journal}
  {\bibinfo  {journal} {Phys. Plasmas}\ }\textbf {\bibinfo {volume} {20}},\
  \bibinfo {pages} {102517} (\bibinfo {year} {2013})}\BibitemShut {NoStop}%
\bibitem [{\citenamefont {Xiao}\ \emph {et~al.}(2015)\citenamefont {Xiao},
  \citenamefont {Liu}, \citenamefont {Qin}, \citenamefont {Yu},\ and\
  \citenamefont {Xiang}}]{xiao2015variational}%
  \BibitemOpen
  \bibfield  {author} {\bibinfo {author} {\bibfnamefont {J.}~\bibnamefont
  {Xiao}}, \bibinfo {author} {\bibfnamefont {J.}~\bibnamefont {Liu}}, \bibinfo
  {author} {\bibfnamefont {H.}~\bibnamefont {Qin}}, \bibinfo {author}
  {\bibfnamefont {Z.}~\bibnamefont {Yu}}, \ and\ \bibinfo {author}
  {\bibfnamefont {N.}~\bibnamefont {Xiang}},\ }\href@noop {} {\bibfield
  {journal} {\bibinfo  {journal} {Physics of Plasmas (1994-present)}\ }\textbf
  {\bibinfo {volume} {22}},\ \bibinfo {pages} {092305} (\bibinfo {year}
  {2015})}\BibitemShut {NoStop}%
\bibitem [{\citenamefont {Evstatiev}\ and\ \citenamefont
  {Shadwick}(2013)}]{evstatiev2013variational}%
  \BibitemOpen
  \bibfield  {author} {\bibinfo {author} {\bibfnamefont {E.}~\bibnamefont
  {Evstatiev}}\ and\ \bibinfo {author} {\bibfnamefont {B.}~\bibnamefont
  {Shadwick}},\ }\href@noop {} {\bibfield  {journal} {\bibinfo  {journal}
  {Journal of Computational Physics}\ }\textbf {\bibinfo {volume} {245}},\
  \bibinfo {pages} {376} (\bibinfo {year} {2013})}\BibitemShut {NoStop}%
\bibitem [{\citenamefont {Evstatiev}(2014)}]{evstatiev2014application}%
  \BibitemOpen
  \bibfield  {author} {\bibinfo {author} {\bibfnamefont {E.}~\bibnamefont
  {Evstatiev}},\ }\href@noop {} {\bibfield  {journal} {\bibinfo  {journal}
  {Computer Physics Communications}\ }\textbf {\bibinfo {volume} {185}},\
  \bibinfo {pages} {2851} (\bibinfo {year} {2014})}\BibitemShut {NoStop}%
\bibitem [{\citenamefont {Shadwick}\ \emph {et~al.}(2014)\citenamefont
  {Shadwick}, \citenamefont {Stamm},\ and\ \citenamefont
  {Evstatiev}}]{Shadwick14}%
  \BibitemOpen
  \bibfield  {author} {\bibinfo {author} {\bibfnamefont {B.~A.}\ \bibnamefont
  {Shadwick}}, \bibinfo {author} {\bibfnamefont {A.~B.}\ \bibnamefont {Stamm}},
  \ and\ \bibinfo {author} {\bibfnamefont {E.~G.}\ \bibnamefont {Evstatiev}},\
  }\href@noop {} {\bibfield  {journal} {\bibinfo  {journal} {Physics of
  Plasmas}\ }\textbf {\bibinfo {volume} {21}},\ \bibinfo {pages} {055708}
  (\bibinfo {year} {2014})}\BibitemShut {NoStop}%
\bibitem [{\citenamefont {Qin}\ \emph {et~al.}(2015{\natexlab{a}})\citenamefont
  {Qin}, \citenamefont {Liu}, \citenamefont {Xiao}, \citenamefont {Zhang},
  \citenamefont {He}, \citenamefont {Wang}, \citenamefont {Burby},
  \citenamefont {Ellison},\ and\ \citenamefont {Zhou}}]{qin2015canonical}%
  \BibitemOpen
  \bibfield  {author} {\bibinfo {author} {\bibfnamefont {H.}~\bibnamefont
  {Qin}}, \bibinfo {author} {\bibfnamefont {J.}~\bibnamefont {Liu}}, \bibinfo
  {author} {\bibfnamefont {J.}~\bibnamefont {Xiao}}, \bibinfo {author}
  {\bibfnamefont {R.}~\bibnamefont {Zhang}}, \bibinfo {author} {\bibfnamefont
  {Y.}~\bibnamefont {He}}, \bibinfo {author} {\bibfnamefont {Y.}~\bibnamefont
  {Wang}}, \bibinfo {author} {\bibfnamefont {J.~W.}\ \bibnamefont {Burby}},
  \bibinfo {author} {\bibfnamefont {L.}~\bibnamefont {Ellison}}, \ and\
  \bibinfo {author} {\bibfnamefont {Y.}~\bibnamefont {Zhou}},\ }\href@noop {}
  {\bibfield  {journal} {\bibinfo  {journal} {arXiv preprint arXiv:1503.08334,
  Nuclear Fusion, in press}\ } (\bibinfo {year}
  {2015}{\natexlab{a}})}\BibitemShut {NoStop}%
\bibitem [{\citenamefont {Crouseilles}\ \emph {et~al.}(2015)\citenamefont
  {Crouseilles}, \citenamefont {Einkemmer},\ and\ \citenamefont
  {Faou}}]{crouseilles2015hamiltonian}%
  \BibitemOpen
  \bibfield  {author} {\bibinfo {author} {\bibfnamefont {N.}~\bibnamefont
  {Crouseilles}}, \bibinfo {author} {\bibfnamefont {L.}~\bibnamefont
  {Einkemmer}}, \ and\ \bibinfo {author} {\bibfnamefont {E.}~\bibnamefont
  {Faou}},\ }\href@noop {} {\bibfield  {journal} {\bibinfo  {journal} {Journal
  of Computational Physics}\ }\textbf {\bibinfo {volume} {283}},\ \bibinfo
  {pages} {224} (\bibinfo {year} {2015})}\BibitemShut {NoStop}%
\bibitem [{\citenamefont {Qin}\ \emph {et~al.}(2015{\natexlab{b}})\citenamefont
  {Qin}, \citenamefont {He}, \citenamefont {Zhang}, \citenamefont {Liu},
  \citenamefont {Xiao},\ and\ \citenamefont {Wang}}]{Qin15JCP}%
  \BibitemOpen
  \bibfield  {author} {\bibinfo {author} {\bibfnamefont {H.}~\bibnamefont
  {Qin}}, \bibinfo {author} {\bibfnamefont {Y.}~\bibnamefont {He}}, \bibinfo
  {author} {\bibfnamefont {R.}~\bibnamefont {Zhang}}, \bibinfo {author}
  {\bibfnamefont {J.}~\bibnamefont {Liu}}, \bibinfo {author} {\bibfnamefont
  {J.}~\bibnamefont {Xiao}}, \ and\ \bibinfo {author} {\bibfnamefont
  {Y.}~\bibnamefont {Wang}},\ }\href {\doibase
  http://dx.doi.org/10.1016/j.jcp.2015.04.056} {\bibfield  {journal} {\bibinfo
  {journal} {Journal of Computational Physics}\ }\textbf {\bibinfo {volume}
  {297}},\ \bibinfo {pages} {721 } (\bibinfo {year}
  {2015}{\natexlab{b}})}\BibitemShut {NoStop}%
\bibitem [{\citenamefont {Morrison}(1980)}]{Morrison80}%
  \BibitemOpen
  \bibfield  {author} {\bibinfo {author} {\bibfnamefont {P.~J.}\ \bibnamefont
  {Morrison}},\ }\href@noop {} {\bibfield  {journal} {\bibinfo  {journal}
  {Physics Letters}\ }\textbf {\bibinfo {volume} {80A}},\ \bibinfo {pages}
  {383} (\bibinfo {year} {1980})}\BibitemShut {NoStop}%
\bibitem [{\citenamefont {Weinstein}\ and\ \citenamefont
  {Morrison}(1981)}]{Weinstein81}%
  \BibitemOpen
  \bibfield  {author} {\bibinfo {author} {\bibfnamefont {A.}~\bibnamefont
  {Weinstein}}\ and\ \bibinfo {author} {\bibfnamefont {P.~J.}\ \bibnamefont
  {Morrison}},\ }\href@noop {} {\bibfield  {journal} {\bibinfo  {journal}
  {Physics Letters}\ }\textbf {\bibinfo {volume} {86A}},\ \bibinfo {pages}
  {235} (\bibinfo {year} {1981})}\BibitemShut {NoStop}%
\bibitem [{\citenamefont {Marsden}\ and\ \citenamefont
  {Weinstein}(1982)}]{marsden1982hamiltonian}%
  \BibitemOpen
  \bibfield  {author} {\bibinfo {author} {\bibfnamefont {J.~E.}\ \bibnamefont
  {Marsden}}\ and\ \bibinfo {author} {\bibfnamefont {A.}~\bibnamefont
  {Weinstein}},\ }\href@noop {} {\bibfield  {journal} {\bibinfo  {journal}
  {Physica D: Nonlinear Phenomena}\ }\textbf {\bibinfo {volume} {4}},\ \bibinfo
  {pages} {394} (\bibinfo {year} {1982})}\BibitemShut {NoStop}%
\bibitem [{\citenamefont {Burby}\ \emph {et~al.}(2015)\citenamefont {Burby},
  \citenamefont {Brizard}, \citenamefont {Morrison},\ and\ \citenamefont
  {Qin}}]{burby2014hamiltonian}%
  \BibitemOpen
  \bibfield  {author} {\bibinfo {author} {\bibfnamefont {J.}~\bibnamefont
  {Burby}}, \bibinfo {author} {\bibfnamefont {A.}~\bibnamefont {Brizard}},
  \bibinfo {author} {\bibfnamefont {P.}~\bibnamefont {Morrison}}, \ and\
  \bibinfo {author} {\bibfnamefont {H.}~\bibnamefont {Qin}},\ }\href {\doibase
  http://dx.doi.org/10.1016/j.physleta.2015.06.051} {\bibfield  {journal}
  {\bibinfo  {journal} {Physics Letters A}\ }\textbf {\bibinfo {volume}
  {379}},\ \bibinfo {pages} {2073 } (\bibinfo {year} {2015})}\BibitemShut
  {NoStop}%
\bibitem [{\citenamefont {Zhou}\ \emph {et~al.}(2014)\citenamefont {Zhou},
  \citenamefont {Qin}, \citenamefont {Burby},\ and\ \citenamefont
  {Bhattacharjee}}]{zhou2014variational}%
  \BibitemOpen
  \bibfield  {author} {\bibinfo {author} {\bibfnamefont {Y.}~\bibnamefont
  {Zhou}}, \bibinfo {author} {\bibfnamefont {H.}~\bibnamefont {Qin}}, \bibinfo
  {author} {\bibfnamefont {J.}~\bibnamefont {Burby}}, \ and\ \bibinfo {author}
  {\bibfnamefont {A.}~\bibnamefont {Bhattacharjee}},\ }\href@noop {} {\bibfield
   {journal} {\bibinfo  {journal} {Physics of Plasmas (1994-present)}\ }\textbf
  {\bibinfo {volume} {21}},\ \bibinfo {pages} {102109} (\bibinfo {year}
  {2014})}\BibitemShut {NoStop}%
\bibitem [{\citenamefont {Zhou}\ \emph {et~al.}(2015)\citenamefont {Zhou},
  \citenamefont {Huang}, \citenamefont {Qin},\ and\ \citenamefont
  {Bhattacharjee}}]{zhou2015formation}%
  \BibitemOpen
  \bibfield  {author} {\bibinfo {author} {\bibfnamefont {Y.}~\bibnamefont
  {Zhou}}, \bibinfo {author} {\bibfnamefont {Y.-M.}\ \bibnamefont {Huang}},
  \bibinfo {author} {\bibfnamefont {H.}~\bibnamefont {Qin}}, \ and\ \bibinfo
  {author} {\bibfnamefont {A.}~\bibnamefont {Bhattacharjee}},\ }\href@noop {}
  {\bibfield  {journal} {\bibinfo  {journal} {arXiv preprint arXiv:1509.08163}\
  } (\bibinfo {year} {2015})}\BibitemShut {NoStop}%
\bibitem [{\citenamefont {Hirani}(2003)}]{hirani2003discrete}%
  \BibitemOpen
  \bibfield  {author} {\bibinfo {author} {\bibfnamefont {A.~N.}\ \bibnamefont
  {Hirani}},\ }\emph {\bibinfo {title} {Discrete Exterior Calculus}},\
  \href@noop {} {Ph.D. thesis},\ \bibinfo  {school} {California Institute of
  Technology} (\bibinfo {year} {2003})\BibitemShut {NoStop}%
\bibitem [{\citenamefont {Stern}\ \emph {et~al.}(2015)\citenamefont {Stern},
  \citenamefont {Tong}, \citenamefont {Desbrun},\ and\ \citenamefont
  {Marsden}}]{stern2007geometric}%
  \BibitemOpen
  \bibfield  {author} {\bibinfo {author} {\bibfnamefont {A.}~\bibnamefont
  {Stern}}, \bibinfo {author} {\bibfnamefont {Y.}~\bibnamefont {Tong}},
  \bibinfo {author} {\bibfnamefont {M.}~\bibnamefont {Desbrun}}, \ and\
  \bibinfo {author} {\bibfnamefont {J.~E.}\ \bibnamefont {Marsden}},\ }in\
  \href@noop {} {\emph {\bibinfo {booktitle} {Geometry, Mechanics, and
  Dynamics}}}\ (\bibinfo  {publisher} {Springer},\ \bibinfo {year} {2015})\
  pp.\ \bibinfo {pages} {437--475}\BibitemShut {NoStop}%
\bibitem [{\citenamefont {Whitney}(1957)}]{whitney1957geometric}%
  \BibitemOpen
  \bibfield  {author} {\bibinfo {author} {\bibfnamefont {H.}~\bibnamefont
  {Whitney}},\ }\href@noop {} {\emph {\bibinfo {title} {Geometric Integration
  Theory}}}\ (\bibinfo  {publisher} {Princeton University Press},\ \bibinfo
  {year} {1957})\BibitemShut {NoStop}%
\bibitem [{\citenamefont {Birdsall}\ and\ \citenamefont
  {Langdon}(1991)}]{birdsall1991plasma}%
  \BibitemOpen
  \bibfield  {author} {\bibinfo {author} {\bibfnamefont {C.~K.}\ \bibnamefont
  {Birdsall}}\ and\ \bibinfo {author} {\bibfnamefont {A.~B.}\ \bibnamefont
  {Langdon}},\ }\href@noop {} {\emph {\bibinfo {title} {Plasma Physics via
  Computer Simulation}}}\ (\bibinfo  {publisher} {IOP Publishing},\ \bibinfo
  {year} {1991})\ p.\ \bibinfo {pages} {293}\BibitemShut {NoStop}%
\bibitem [{\citenamefont {Hockney}\ and\ \citenamefont
  {Eastwood}(1988)}]{hockney1988computer}%
  \BibitemOpen
  \bibfield  {author} {\bibinfo {author} {\bibfnamefont {R.~W.}\ \bibnamefont
  {Hockney}}\ and\ \bibinfo {author} {\bibfnamefont {J.~W.}\ \bibnamefont
  {Eastwood}},\ }\href@noop {} {\emph {\bibinfo {title} {Computer Simulation
  Using Particles}}}\ (\bibinfo  {publisher} {CRC Press},\ \bibinfo {year}
  {1988})\BibitemShut {NoStop}%
\bibitem [{\citenamefont {Nieter}\ and\ \citenamefont
  {Cary}(2004)}]{nieter2004vorpal}%
  \BibitemOpen
  \bibfield  {author} {\bibinfo {author} {\bibfnamefont {C.}~\bibnamefont
  {Nieter}}\ and\ \bibinfo {author} {\bibfnamefont {J.~R.}\ \bibnamefont
  {Cary}},\ }\href@noop {} {\bibfield  {journal} {\bibinfo  {journal} {Journal
  of Computational Physics}\ }\textbf {\bibinfo {volume} {196}},\ \bibinfo
  {pages} {448} (\bibinfo {year} {2004})}\BibitemShut {NoStop}%
\bibitem [{\citenamefont {Qin}(2005)}]{QinFields}%
  \BibitemOpen
  \bibfield  {author} {\bibinfo {author} {\bibfnamefont {H.}~\bibnamefont
  {Qin}},\ }\href@noop {} {\bibfield  {journal} {\bibinfo  {journal} {Fields
  Institute Communications}\ }\textbf {\bibinfo {volume} {46}},\ \bibinfo
  {pages} {171} (\bibinfo {year} {2005})}\BibitemShut {NoStop}%
\bibitem [{\citenamefont {Crouseilles}\ \emph {et~al.}(2007)\citenamefont
  {Crouseilles}, \citenamefont {Mehrenberger},\ and\ \citenamefont
  {Sonnendrucker}}]{Crouseilles07}%
  \BibitemOpen
  \bibfield  {author} {\bibinfo {author} {\bibfnamefont {N.}~\bibnamefont
  {Crouseilles}}, \bibinfo {author} {\bibfnamefont {M.}~\bibnamefont
  {Mehrenberger}}, \ and\ \bibinfo {author} {\bibfnamefont {E.}~\bibnamefont
  {Sonnendrucker}},\ }\href@noop {} {\bibfield  {journal} {\bibinfo  {journal}
  {Journal of Computational Physics}\ }\textbf {\bibinfo {volume} {229}},\
  \bibinfo {pages} {1927} (\bibinfo {year} {2007})}\BibitemShut {NoStop}%
\bibitem [{\citenamefont {Dawson}(1961)}]{dawson1961landau}%
  \BibitemOpen
  \bibfield  {author} {\bibinfo {author} {\bibfnamefont {J.}~\bibnamefont
  {Dawson}},\ }\href@noop {} {\bibfield  {journal} {\bibinfo  {journal}
  {Physics of Fluids (1958-1988)}\ }\textbf {\bibinfo {volume} {4}},\ \bibinfo
  {pages} {869} (\bibinfo {year} {1961})}\BibitemShut {NoStop}%
\bibitem [{\citenamefont {O'Neil}(1965)}]{o1965collisionless}%
  \BibitemOpen
  \bibfield  {author} {\bibinfo {author} {\bibfnamefont {T.}~\bibnamefont
  {O'Neil}},\ }\href@noop {} {\bibfield  {journal} {\bibinfo  {journal}
  {Physics of Fluids (1958-1988)}\ }\textbf {\bibinfo {volume} {8}},\ \bibinfo
  {pages} {2255} (\bibinfo {year} {1965})}\BibitemShut {NoStop}%
\bibitem [{\citenamefont {Mouhot}\ and\ \citenamefont
  {Villani}(2011)}]{mouhot2011landau}%
  \BibitemOpen
  \bibfield  {author} {\bibinfo {author} {\bibfnamefont {C.}~\bibnamefont
  {Mouhot}}\ and\ \bibinfo {author} {\bibfnamefont {C.}~\bibnamefont
  {Villani}},\ }\href@noop {} {\bibfield  {journal} {\bibinfo  {journal} {Acta
  Mathematica}\ }\textbf {\bibinfo {volume} {207}},\ \bibinfo {pages} {29}
  (\bibinfo {year} {2011})}\BibitemShut {NoStop}%
\bibitem [{\citenamefont {Manfredi}(1997)}]{manfredi1997long}%
  \BibitemOpen
  \bibfield  {author} {\bibinfo {author} {\bibfnamefont {G.}~\bibnamefont
  {Manfredi}},\ }\href@noop {} {\bibfield  {journal} {\bibinfo  {journal}
  {Physical Review Letters}\ }\textbf {\bibinfo {volume} {79}},\ \bibinfo
  {pages} {2815} (\bibinfo {year} {1997})}\BibitemShut {NoStop}%
\bibitem [{\citenamefont {Zhou}\ \emph {et~al.}(2001)\citenamefont {Zhou},
  \citenamefont {Guo},\ and\ \citenamefont {Shu}}]{zhou2001numerical}%
  \BibitemOpen
  \bibfield  {author} {\bibinfo {author} {\bibfnamefont {T.}~\bibnamefont
  {Zhou}}, \bibinfo {author} {\bibfnamefont {Y.}~\bibnamefont {Guo}}, \ and\
  \bibinfo {author} {\bibfnamefont {C.-W.}\ \bibnamefont {Shu}},\ }\href@noop
  {} {\bibfield  {journal} {\bibinfo  {journal} {Physica D: Nonlinear
  Phenomena}\ }\textbf {\bibinfo {volume} {157}},\ \bibinfo {pages} {322}
  (\bibinfo {year} {2001})}\BibitemShut {NoStop}%
\bibitem [{\citenamefont {Stix}(1992)}]{stix1992waves}%
  \BibitemOpen
  \bibfield  {author} {\bibinfo {author} {\bibfnamefont {T.~H.}\ \bibnamefont
  {Stix}},\ }\href@noop {} {\emph {\bibinfo {title} {Waves in Plasmas}}}\
  (\bibinfo  {publisher} {Springer},\ \bibinfo {year} {1992})\ pp.\ \bibinfo
  {pages} {276--277}\BibitemShut {NoStop}%
\bibitem [{\citenamefont {Desbrun}\ \emph {et~al.}(2008)\citenamefont
  {Desbrun}, \citenamefont {Kanso},\ and\ \citenamefont
  {Tong}}]{desbrun2008discrete}%
  \BibitemOpen
  \bibfield  {author} {\bibinfo {author} {\bibfnamefont {M.}~\bibnamefont
  {Desbrun}}, \bibinfo {author} {\bibfnamefont {E.}~\bibnamefont {Kanso}}, \
  and\ \bibinfo {author} {\bibfnamefont {Y.}~\bibnamefont {Tong}},\ }in\
  \href@noop {} {\emph {\bibinfo {booktitle} {Discrete Differential
  Geometry}}}\ (\bibinfo  {publisher} {Springer},\ \bibinfo {year} {2008})\
  pp.\ \bibinfo {pages} {287--324}\BibitemShut {NoStop}%
\end{thebibliography}%

\end{document}